\documentclass[a4paper,fleqn,usenatbib]{mnras}
\usepackage{newtxtext,newtxmath}
\usepackage[T1]{fontenc}
\usepackage{ae,aecompl}
\usepackage{multicol}
\usepackage{graphicx}	
\usepackage{amsmath}	

\usepackage{cleveref}
\usepackage{pdflscape}

\newcommand{\kms}{{\mathrm{km~s^{-1}}}}

\title[Asteroseismology of AE UMa and RV Ari]{Asteroseismology of
double-mode radial $\delta$ Scuti stars: \\AE Ursae Majoris and RV Arietis}

\author[Daszy\'nska-Daszkiewicz et al.]{ J. Daszy\'nska-Daszkiewicz\thanks{E-mail:daszynska@astro.uni.wroc.pl},
P. Walczak, W. Szewczuk, W. Niewiadomski\\
Instytut Astronomiczny, Uniwersytet Wroc{\l}awski, Kopernika 11, 51-622 Wroc{\l}aw, Poland\\
}

\date{Accepted XXX. Received YYY; in original form ZZZ}

\pubyear{2021}

\hypersetup{draft}
\begin{document}
\label{firstpage}
\pagerange{\pageref{firstpage}--\pageref{lastpage}}
\maketitle

\begin{abstract}
We construct complex seismic models of two high-amplitude $\delta$ Sct stars, AE UMa and RV Ari,
each pulsating in two radial modes: fundamental and first overtone.
The models reproduce, besides the frequencies of two radial modes, also the amplitude of bolometric flux variations
(the non-adiabatic parameter $f$) for the dominant mode.
Applying the Monte Carlo–based Bayesian analysis, we derive strong constraints, on the parameters of the model
as well as on the free parameters of the theory.
A vast majority of seismic models of the two stars are just at the beginning of hydrogen-shell burning and a small
fraction is at the very end of an overall contraction.
The stars have a similar age of about 1.6\,Gyr for the hydrogen-shell burning phase.
Both stars have unusual low overshooting from the convective core;  about 0.02 and 0.004 of the pressure scale height
for AE UMa and RV Ari, respectively. This result presumably indicates that overshooting should vary with time 
and scale with a decreasing convective core.
The efficiency of convection in the envelope of both stars is rather low and described by the mixing length parameter 
$\alpha_{\rm MLT}$ of about 0.3$-$0.6.
The third frequency of RV Ari, confirmed by us in the TESS photometry, can only be associated with mixed nonradial 
modes $\ell=1,~g_{4}-g_{8}$ or $\ell=2,~g_{10}-g_{12}$.
We include the dipole mode into our Bayesian modelling and demonstrate its huge asteroseismic potential.
\end{abstract}



\section{Introduction}
Asteroseismology of stars pulsating in more than one radial mode is of particular importance
because the period ratio of such modes takes values in a very narrow range.
The high-amplitude $\delta$ Scuti stars (HADS), a special subclass of $\delta$ Sct variables,
often pulsate in two radial modes, usually in the fundamental and first overtone mode
\citep[e.g.,][]{Breger2000, McNamara2000, Furgoni2016, Yang2021}.
$\delta$ Scuti stars are classical pulsating variables of AF spectral type and their instability is driven
by the opacity mechanism operating in the second helium ionization zone  \citep{Chevalier1971}, with
a small contribution from the hydrogen ionization region \citep{Pamyatnykh1999}.
The masses of $\delta$ Sct pulsators are in the range of about 1.6 - 2.6 $\mathrm{M}_\odot$
and most of them are in the main-sequence phase of evolution \citep[e.g.,][]{BregerPam1998, Bowman2016}.
Radial and non-radial pulsations in pressure (p) and gravity (g) modes are be excited.

HADS stars change their brightness in the $V$-passband in the range greater than 0.3\,mag.
They are in an advanced phase of main-sequence evolution or, usually, already in a post-main sequence phase
\citep[e.g.,][]{Breger2000}. HADS pulsators have typically low rotational velocities, below $V_{\rm rot}\sin i = 40~\kms$ \citep{Breger2000}, 
although there is at least one exception, i.e., V2367 Cyg with the rotational velocity of about $100~\kms$ \citep{Balona2012}.

From the fitting of frequencies of just two radial modes, one can already obtain valuable constraints
on global stellar parameters such as a mass, effective temperature, luminosity  \citep[e.g.,][]{Petersen1996,JDD2022,Netzel2022}. 
However, to get more unambiguous solution and  more information about a star, e.g., on chemical composition, 
mixing processes or efficiency of convection, including nonradial modes or other seismic tools is essential.
In particular, the non-adiabatic parameter $f$ is most suitable
for obtaining reliable constraints on convection  in the outer layers of $\delta$ Sct stars \citep{JDD2003}.
The parameter $f$ gives the relative amplitude of the radiative flux perturbation at the photosphereic level.
Its diagnostic potential for constraining the mixing length parameter $\alpha_{\rm MLT}$  has been already demonstrated 
many times for the AF-type pulsators, e.g., $\beta$ Cas, AB Cas, 20 CVn \citep{JDD2003, JDD2007},
FG Vir \citep{JDD2005},  SX Phe \citep{JDD2020,JDD2023}, the prototype $\delta$ Sct \citep{JDD2021} and
BP Peg  \citep{JDD2022,JDD2023}.  The main results for AE UMa and RV Ari  were published by \citet{JDD2023}, 
where we showed for the four HADS stars that  only the seismic models computed with the OPAL opacities \citep{Iglesias1996}
are caught within the observed error box in the HR diagram. Seismic models computed with OP tables \citep{Seaton2005} 
and OPLIB tables \citep{Colgan2016} were much cooler and less luminous.

Here, we present the details of complex seismic modelling of AE UMa and RV Ari,  which relies on the simultaneous fitting
of the two radial modes  and the non-adiabatic parameter $f$ for the dominant mode. Besides, we present the Fourier  frequency analysis 
of  the TESS  space data,  mode identification from the phototometric observables and the asteroseismic potential 
of the nonradial mode present in the star RV Ari.

Sect.\,2 contains basic information about the stars and determination of main observational parameters. 
In Sec.\,3, we present the frequency analysis 
of the TESS data of the two HADS pulsators. In the case of RV Ari, the ASAS photometry is also analysed.
In Sect.\,4, we identify the degree $\ell$ of two pulsational modes using the method based 
on the photometric amplitudes and phases, to confirm, independently of the period ratio, their radial nature.
Sect.\,5, presents the details of our complex seismic modelling of both HADS stars
based on the Bayesian analysis using Monte Carlo simulations. In Sect.\,6 we include the nonradial mode into seismic modelling of RV Ari. The summary is given in Sect.\,7.

\section{The two double-mode radial pulsators: AE UMa and RV Ari}
\begin{figure*}
	\includegraphics[width=175mm,clip]{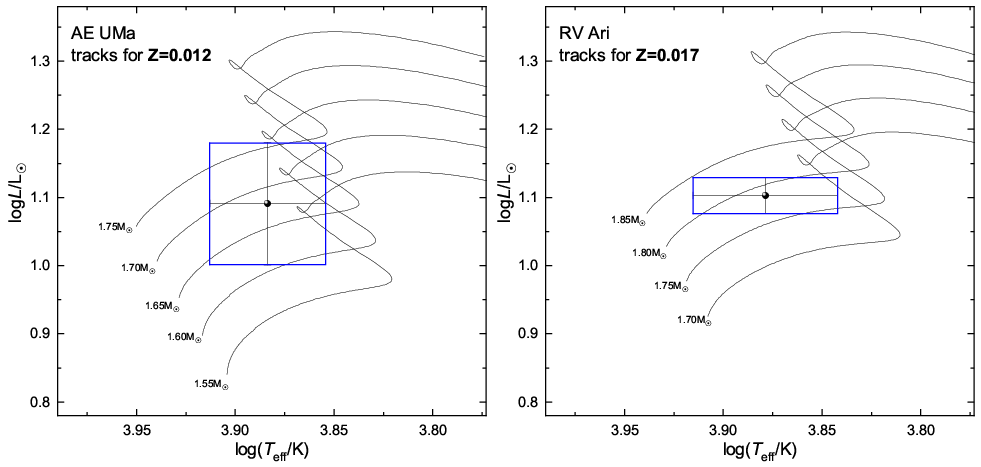}
	\caption{The HR diagrams with the position of AE UMa (left panel) and RV Ari (right panel).
                The evolutionary tracks were computed with Warsaw-New Jersey code, assuming the OPAL opacity tables, the AGSS09 solar mixture and the initial hydrogen  abundance $X_0=0.70$.  The metallicity is indicated in each panel.  The mixing length
                parameter was $\alpha_{\rm MLT}=0.5$ and zero-overshooting from the convective core was adopted. The initial velocity 
                of rotation was $V_{\rm rot,0}=10~\kms$.}
	\label{fig1}
\end{figure*}
AE Ursae Majoris is an A9-spectral type star  with the mean brightness in the V passband of 11.35\,mag  (SIMBAD Astronomical Database).
The variability of the star was discovered by \citet{Greyer1955} and, firstly, is was classified as a dwarf Cepheid 
by \citet{Tsesevich1973} who determined the period of light variations. The secondary period was found by \citet{Szeidl1974}
and \citet{Broglia1975}.
\citet{Garcia1995} listed it as an SX Phe variable and this classification is still in GCVS and on the SIMBAD website.
However, already \citet{Cox1979} postulated, on the basis of the period ratio, that AE UMa is a Population I high-amplitude 
$\delta$ Sct star. Moreover, \citet{Rodriguez1992} showed that the metallicty of AE UMa is [m/H]$=-0.3$, using the photometric 
index $\delta m_1$. \citet{Hintz1997} determined  [m/H] from $-0.1$ to $-0.4$ using an approximate relationship 
between the metallicity and the period ratio. Thus, there is no doubt that AE UMa belongs to Population I.
As a majority of HADS stars, AE UMa is a slow rotator with $V\sin i \approx 28~\kms$ \citep{Jonsson2020}.
\citet{Pocs2001}  analysing  25 years of photometric observations concluded that the period of the fundamental radial mode 
is stable and the period of the first overtone is decreasing with a rate ${\dot P}/P=-7.3\cdot 10^{-8}$\,yr$^{-1}$.
According to the authors amplitudes of both modes undergo only small changes.

\citet{Niu2017} constructed for the first time seismic models of  AE UMa based on the two radial modes and the period changes.
From about 440 times of maximum light, they determined the positive period change for the dominant mode 
with a rate of ${\dot P}/P=+5.4(1.9)\cdot 10^{-9}$\,yr$^{-1}$.   
In their seismic modelling of AE UMa, \citet{Niu2017} ignored  all effects of rotation and fixed  the values of overshooting parameter 
from a convective core $f_{\rm ov}=0.015$ and the mixing length parameter $\alpha_{\rm MLT}=1.89$.   
They concluded that AE UMa is in the post-MS stage of evolution. 
Recently,  \cite{2022RAA....22j5006X} performed the frequency analysis of  the TESS data of AE UMa made in sector 21.
They found two independent frequencies, $11.6257(2)\,\mathrm{d}^{-1}$ and $15.0314(2)\,\mathrm{d}^{-1}$, 
as well as 63 harmonics and combinations of them. Using the times of maximum light from about 46 years,  they obtained 
${\dot P}/P=+2.96(5)\cdot 10^{-9}$\,yr$^{-1}$ for the dominant period. \cite{2022RAA....22j5006X}  demonstrated also 
a prospect of using the period changes in asteroseismic modelling and constructed such seismic models for the fixed values of 
the mixing length parameter of $\alpha_{\rm MLT}=1.89$. The authors ignored all effects of rotation 
and assumed zero-overshooting  from the convective core.

RV Arietis is the Population I star with an A spectral type and the mean brightness in the V passband of 12.27\,mag.
The star was identified as variable by \citet{Hoffmeister1934}. \citet{Boglia1955} and \citet{Detre1956} derived the main period 
and detected the second mode from the modulation period. These two periodic variations are explained by excitation
of the fundamental and first overtone radial modes \citep{Cox1979}.
RV Ari is a slow rotator with the projected rotational velocity of $V\sin i \approx 18~\kms$ \citep{Rodriguez2000}.
Using the photometric index $\delta m_1$, \citet{Rodriguez1992} determined for RV Ari the above-solar metallicity of [m/H]$=+0.1$.
\citet{Pocs2002} gathered an extensive BVRI photometry, covering about 20 years, and obtained a decreasing period 
of the fundamental mode with a rate ${\dot P}/P=-0.6\cdot 10^{-8}$\,yr$^{-1}$ and an increasing period for the first overtone
${\dot P}/P=+0.9\cdot 10^{-8}$\,yr$^{-1}$. The opposite sign of period changes for the two modes indicates some non-evolutionary effects.
The authors detected also the third signal in their photometry with the frequency 13.6116\,d$^{-1}$ that can correspond only to nonradial mode.

\citet{Casas2006} presented the first seismic modelling of the star adopting four discrete values of the mixing length parameter 
$\alpha_{\rm MLT}=0.5, 1.0, 1.5, 2.0$. They considered only main-sequence models and obtained the constraints on effective temperature
 [7065,\,7245]\,K and on age  [1.19,\,1.27]\,Gyr.

In our seismic analysis of both stars, we adopted the whole range of the effective temperature found in the literature.
To derive the luminosity, we adopted distances determined on the basis of Starhorse2 model (Anders et al., 2022), 
using the Gaia EDR3 observations (Gaia Collaboration et al., 2022).  The bolometric corrections were taken from Kurucz models
for the microturbulent velocity $\xi_t=2$ and $4\,\kms$. We considered the metallicity [m/H]$=-0.5,~ -0.3,~-0.2,~-0.1$ for AE UMa 
and [m/H]$=0.0,~ +0.1,~+0.3,~+0.5$ for RV Ari. The adopted parameters were as follows:\\
$\bullet$ AE\,UMa:  $\log T_{\rm eff} = 3.88353(2922)$,  $\log L/$L$_\odot  = 1.0907(896)$,\\ 
$\bullet$ ~RV\,Ari:  $\log T_{\rm eff} = 3.8787(367)$, $\log L/$L$_\odot  =1.1029 (263)$.
 
 In Fig.\,1, we show the position of both stars on the Hertzsprung-Russell diagram. As one can see, the stars occupy a similar position
 but their metallicity is quite different; AE UMa has [m/H] below the solar value and RV Ari - above the solar value. 
 For guidance, we show also a few evolutionary tracks computed with  Warsaw-New Jersey code described in Sect.\,5.
 The tracks were computed for the OPAL opacity table \citep{Iglesias1996} and the solar chemical mixture of\citet{Asplund2009},
 hereafter AGSS09.
 
\section{Frequency Analysis}
Both stars, AE UMa and RV Ari, were observed in the framework of the TESS mission \citep{2015JATIS...1a4003R}.
Here, we used corrected 120\,s cadence observations
delivered by TESS Science Processing Operations Center \citep[SPOC,][]{2016SPIE.9913E..3EJ}.

AE UMa was observed in the two 27\,d sectors, S21 and S48, which are more than 2 years apart.
Therefore, we decided to analyse each sector separately. 
The Rayleigh resolution for each sector is about  $\Delta \nu_R=1/T=0.037\,\mathrm{d}^{-1}$.

RV Ari was observed in two sectors, 42 and 43, which span 51 days.
Since these sector are consecutive, we analyze them together. The Rayleigh resolution for the combined sectors
is $0.020\,\mathrm{d}^{-1}$. In addition to the space TESS data, we analysed the ground-based
ASAS-3 V-band photometry \citep[][]{2002AcA....52..397P} of RV Ari.
ASAS data cover 2518 days what translates into the Rayleigh resolution  $\Delta \nu_R=0.0004\,\mathrm{d}^{-1}$).

In the first step, we normalized the TESS light curves by dividing them by a linear fit. Only data points with quality flag 0 were used. 
The normalization was done for each sector separately. In order to extract frequencies of the light variability,
we proceeded the standard pre-whitening procedure.  Amplitude periodograms \citep{1975Ap&SS..36..137D, 1985MNRAS.213..773K}
were calculated up to the Nyquist frequency for TESS 120\,s cadence data, i.e., to 360\,d$^{-1}$. The fixed frequency step 
in periodogram equal to $5\times 10^{-5}\,\mathrm{d}^{-1}$ was used for both analyzed stars.
In the case of TESS data, as a significance criterion of a given frequency peak we chose the signal-to-noise ratio, $S/N=5$.
This threshold is higher than the standard value of 4 \citep{1993spct.conf..106B, 1997A&A...328..544K},
but it corresponds to an estimate made by \citet{2021AcA....71..113B} for TESS data.
The noise $N$ was calculated as the mean value in a one day window centred at the frequency before its extraction.

In the case of data with a high point-to-point precision, there is a risk of artificially introducing
spurious signals in the pre-whitening process. According to \cite{1978Ap&SS..56..285L}, in their most conservative case,
frequencies that are separated less than 2.5 times the Rayleigh resolution cannot be resolved properly and may be spurious.
Therefore, we decided to skip frequency with smaller amplitude in such close pairs.

Adopting the above criteria, in the case of AE UMa we found 59 significant frequency peaks in the S21 data and 57 in the S48 data.
In the case of each sector, two peaks were rejected because of the adopted frequency resolution $2.5\Delta\nu_R$.
For RV Ari, we found 137 frequency peaks. Two frequencies were rejected because of the adopted resolution.

\begin{figure}
	\includegraphics[angle=270, width=\columnwidth]{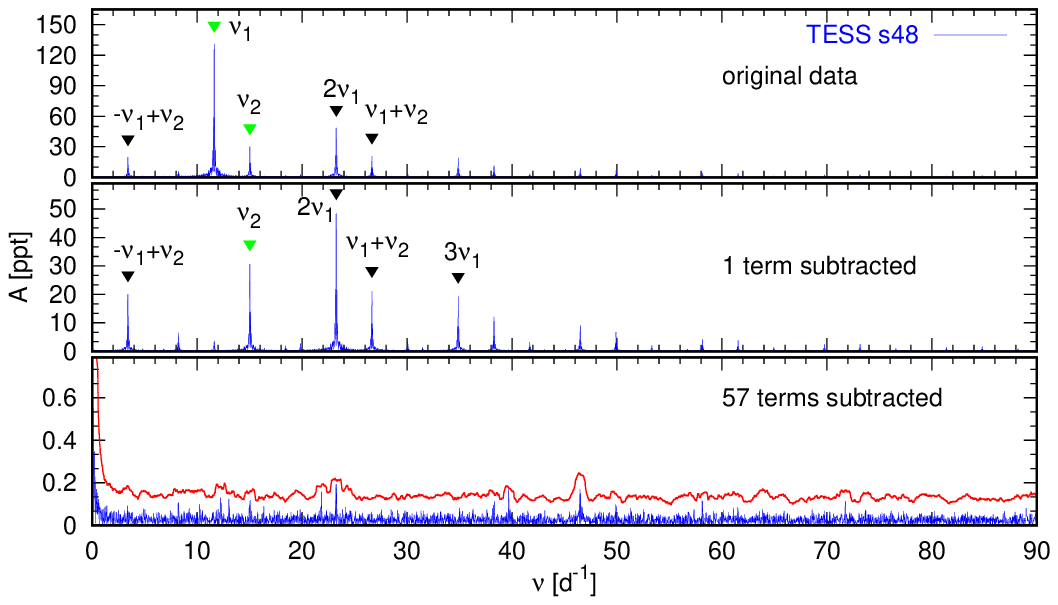}
	\caption{The Fourier amplitude periodograms for AE UMa obtained from the TESS light curve for the sector S48.
		The panels from top to bottom show: the periodogram for the original data, after subtracting one term and after subtracting 57 terms. A red line indicates the $5S/N$ level.}
	\label{fig:AE_UMa_freq_analysis}
\end{figure}

\begin{figure}
	\includegraphics[angle=270, width=\columnwidth]{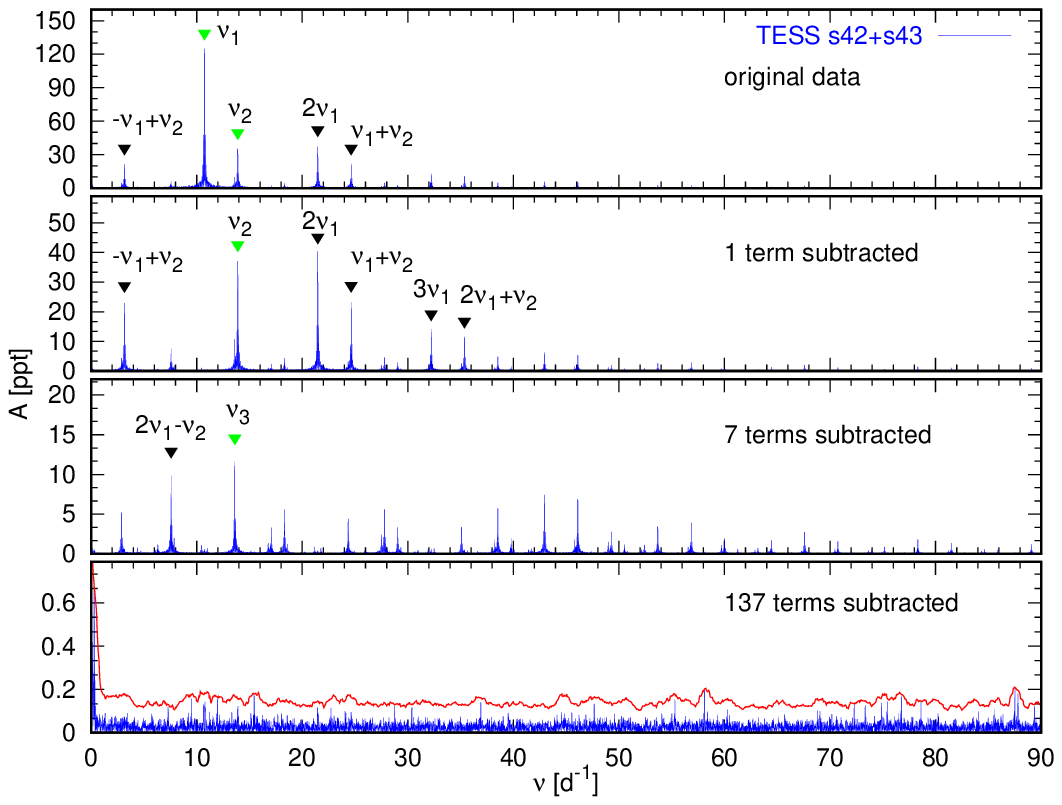}
	\caption{The Fourier amplitude periodograms for RV Ari obtained from the TESS light curve from combined sectors S42 and S43.
		The panels from top to bottom show: the periodogram for the original data, after subtracting one term, after subtracting
		7 terms and after subtracting 137 terms. A red  line indicates the $5S/N$ level.}
	\label{fig:RV_ARi_freq_TESS}
\end{figure}

In Fig.\,\ref{fig:AE_UMa_freq_analysis}, we show the three amplitude periodograms calculated
for the S48 data of AE UMa, i.e, 1)  for the original data (top panel), 2) after subtracting the frequency  $\nu_1$ (middle panel)
and 3) after subtracting all significant frequencies (bottom panel). Periodograms for the S21 data are visually indistinguishable.
In Fig.\,\ref{fig:RV_ARi_freq_TESS}, we show four amplitude periodograms for the S42+S43 data of RV Ari. From top to bottom 
these are: 1) the periodogram calculated  for the original data, 2) after subtracting $\nu_1$,  3) after subtracting $\nu_1$, $\nu_2$ 
and five combinations/harmonics, and 4) after subtracting all significant signals.

\begin{figure}
	\includegraphics[angle=270, width=\columnwidth]{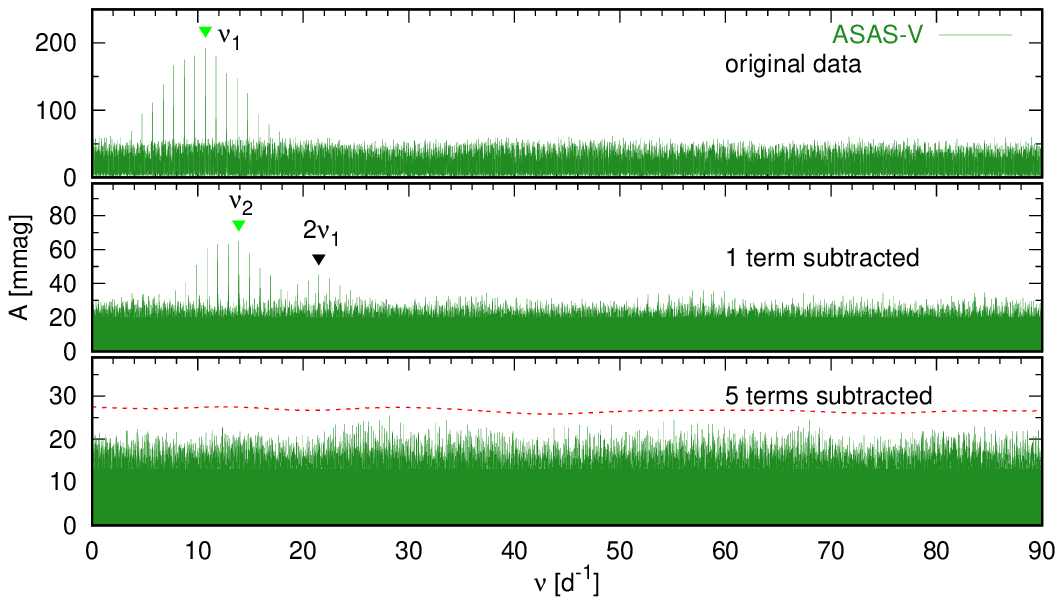}
	\caption{The Fourier amplitude periodograms for RV Ari obtained from the ASAS photometry. The panels from top to bottom show:
		the periodogram for the original data, after subtracting one term and after subtracting 5 terms.
		A red line indicates the $4S/N$ level.}
	\label{fig:RV_ARi_freq_ASAS}
\end{figure}

Our final frequencies, amplitudes and phases were determined using the nonlinear least-squares fit
using the following formula:
\begin{equation}
	S(t)= \sum_{i=1}^N A_i \sin \left( 2\pi \left( \nu_i t + \phi_i \right)  \right) +c,
\end{equation}
where $N$ is the number of sinusoidal components, $A_i$, $\nu_i$, $\phi_i$ are
the amplitude, frequency and phase of the $i-$th component, respectively, while the $c$ is an  offset.
Moreover, we applied the correction to formal frequency errors as suggested by
\citet[][the post-mortem analysis]{1991MNRAS.253..198S}. These corrections for both analyzed stars were of about $1.5$.

Finally, the entire set of frequencies was searched for harmonics and combinations.
A given frequency was considered a combination if it satisfied the equation
\begin{equation}
	\nu_i=m\nu_j+n\nu_k + o \nu_l
	\label{eq:combination}
\end{equation}
within the Rayleigh resolution. In the case of two-parent combinations one of the integers, $m$, $n$ or $o$, was set to zero,
while in the case of harmonics two integers were set to zero.
Moreover, we assumed that $\nu_j$, $\nu_k$ and $\nu_l$ have higher amplitudes than $\nu_i$.

Out of all significant frequency peaks found in TESS data of AE UMa (see Appendix A, Table
\ref{tab:full_TESS_AE_UMa_s21}), only two well known frequencies are independent. We give their values with 
the amplitudes and S/N ratios in Table\,\ref{tab:TESS_AE_UMa}.
Our results agree with the recent analysis of the TESS S21 data
by \cite{2022RAA....22j5006X}. The authors found two independent frequencies, $11.6257(2)\,\mathrm{d}^{-1}$
and $15.0314(2)\,\mathrm{d}^{-1}$ as well as similar to ours harmonics and combinations.

The time span between two sectors of AE UMa is about 740\,d. The difference in a dominant frequency $\nu_1$ obtained from S21 
and S48 is $\Delta\nu=\nu_1(S48)-\nu_1(S21)= -0.000164(7)$\,d$^{-1}$. The corresponding difference in period amounts 
to $\Delta P_1=P_1(48)-P_1(21)=+0.000001214(52)$\,d.
It gives the period change of $\Delta P_1/\Delta t=+5.98(26)\cdot 10^{-7}$\,[d/yr] and a rate of period changes
is ${\dot P_1}/P_1=+6.95(29)\cdot 10^{-6}$\,[yr$^{-1}$]. This rate is about three orders of magnitude higher
than the values obtained by \citet{Niu2017} and \cite{2022RAA....22j5006X} who applied the O-C method to observations
from about 40 years. Such an abrupt  change of period can have a local in time character and results from 
nonlinear interactions of pulsational modes. The rapid period changes over shorter time scales are observed 
for many $\delta$ Scuti stars \citep[e.g.,][]{BregerPam1998,Bowman2021}.
For the first overtone mode we got also an increasing period with the rate of  ${\dot P}_2/P_2=+8.54(99)\cdot 10^{-6}$\,[yr$^{-1}$].
\begin{table}
	\centering
	\caption{Independent frequencies found in the TESS observations of AE UMa from sectors 21 and 48.
		The following columns give the frequency ID, the frequency value, amplitude and signal-to-noise ratio.
		Formal errors are given in parentheses.}
	\label{tab:TESS_AE_UMa}
	\begin{tabular}{ r r r r   } 
		\hline
		ID    &  $\nu\,[\mathrm{d}^{-1}]$               &        A [ppt]         &   $S/N$     \\
        \hline		
		\multicolumn{3}{l}{S21} & \\
		\hline
		$\nu_1$  &     11.625687(3)  &          131.88(2)  &       9.5  \\
		$\nu_2$  &     15.03139(1)   &           30.64(2)  &       9.4  \\
		\hline
		\multicolumn{3}{l}{S48} & \\
        \hline		
		$\nu_1$  &     11.625523(4)  &          131.49(3)  &       8.4  \\
        $\nu_2$  &     15.03113(2)   &           30.77(3)  &       8.4  \\
        \hline
	\end{tabular}
\end{table}
\begin{table}
	\centering
	\caption{Independent frequencies found in the TESS observations of RV Ari from combined sectors 42 and 43.
		The following columns give the frequency ID, frequency value, amplitude and signal-to-noise ratio.
		Formal errors are given in parentheses. }
	\label{tab:TESS_RV_Ari}
	\begin{tabular}{r r r r  } 
		\hline
		ID    &  $\nu\,[\mathrm{d}^{-1}]$               &        A [ppt]     &   $S/N$ \\
		\hline
		$\nu_1$  &       10.737880(3)  &          128.84(2)  &                   11.8  \\
		$\nu_2$  &       13.899137(9)  &           39.63(2)  &                   10.5  \\
		$\nu_3$  &        13.61183(3)  &           11.61(2)  &                   11.9  \\
		\hline
	\end{tabular}
\end{table}
\begin{table}
	\centering
	\caption{The frequencies found in the ASAS data for RV Ari, with the amplitudes, signal-to-noise ratios and comments 
		on possible combinations or harmonics. Errors are given in parenthesis.}
	\label{tab:ASAS_RV_Ari}
	\begin{tabular}{r r r r r  } 
		\hline
		ID    &  $\nu\,[\mathrm{d}^{-1}]$               &        A [mag]    &      $S/N$       & remarks \\
		\hline
		1 &  10.737895(6)   &    0.207(5)  &          10.8  &           \\
		2 &  13.89918(2)    &    0.063(5)  &          6.5   &           \\
		3 &  21.47576(2)    &    0.058(6)  &          5.2   & $2\nu_1$  \\
		4  &  3.16123(3)    &    0.035(5)     &       4.7   & $-1\nu_{1}+1\nu_{3}$ \\
		5  &  24.63706(3)   &     0.035(5)    &        4.7  &  $1\nu_{1}+1\nu_{3}$ \\
		\hline
	\end{tabular}
\end{table}

In the case of RV Ari, three frequencies appeared to be independent. Their values, amplitudes and S/N ratios 
are given in Table\,\ref{tab:TESS_RV_Ari}. 
The rest of 132 peaks can be explained by various combinations of these three independent frequencies.
The third frequency $\nu_3$ has been already suggested from ground-based photometry
by \citet{Pocs2002} and we confirm it in the space data. Thus, RV Ari is one of a few HADS stars with 
an unquestionably existing third independent frequency that can only be associated with a nonradial mode.

Next, we analyzed the ASAS-3 V-band photometry of RV Ari.  These data consist of the photometry made in five different apertures.
We used the one with the smallest mean error. Only points with quality flag A and B were retained. In the case of ground-based photometry,
 we adopted $S/N=4$ as a threshold for significant frequency peaks. The noise was calculated in a wider window of 10\,d$^{-1}$. 
The amplitude periodograms were calculated up to $150\,\mathrm{d}^{-1}$, what covers frequency range found in TESS data.
The step in periodograms was set to $10^{-5}\,\mathrm{d}^{-1}$.  We found five significant signals, two of them are independent
and the remaining three are combinations and harmonic. These signals with the amplitudes and $S/N$ ratios are given in Table\,\ref{tab:ASAS_RV_Ari}.
In Fig.\,\ref{fig:RV_ARi_freq_ASAS}, we show the amplitude periodograms calculated for original data (top panel), 
after subtracting one frequency (middle panel) and after subtracting all 5 significant frequencies (bottom panel).

\section{Identification of the mode degree $\ell$}
\begin{figure*}
	\includegraphics[width=175mm,clip]{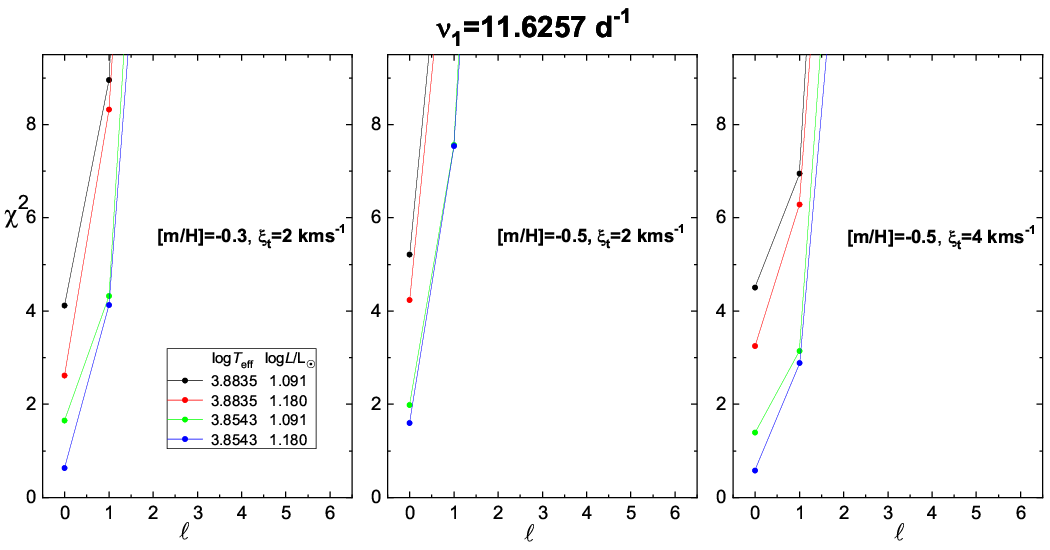}
	\includegraphics[width=175mm,clip]{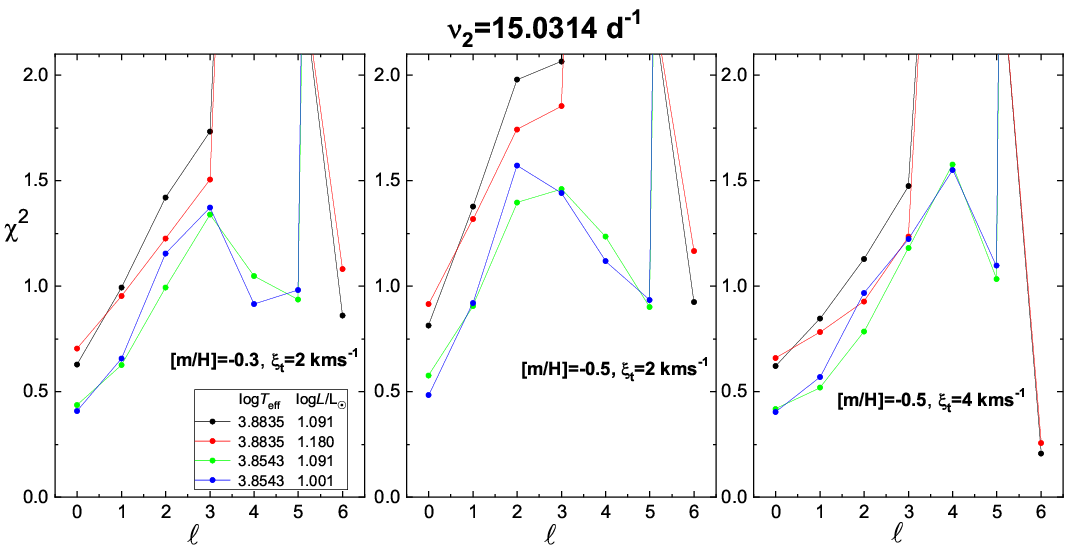}
	\caption{Identification of the mode degree $\ell$ for the frequencies $\nu_1=11.6257$\,d$^{-1}$ (upper panels) and
		$\nu_2=15.0314$\,d$^{-1}$ (lower panels) of AE UMa. The values of $\log T_{\rm eff}$ and $\log L/{\rm L}_{\sun}$
		include the observed error box. Three combinations of the atmospheric metallicity [m/H] and microturbulent velocity $\xi_t$
		were considered.}
	\label{fig5}
\end{figure*}

\begin{figure*}
	\includegraphics[width=175mm,clip]{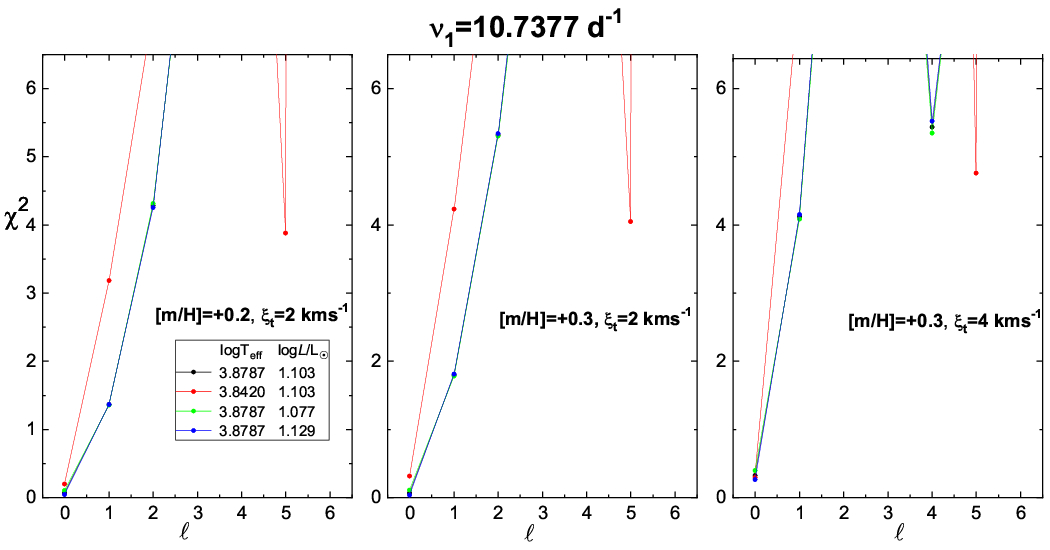}
	\includegraphics[width=175mm,clip]{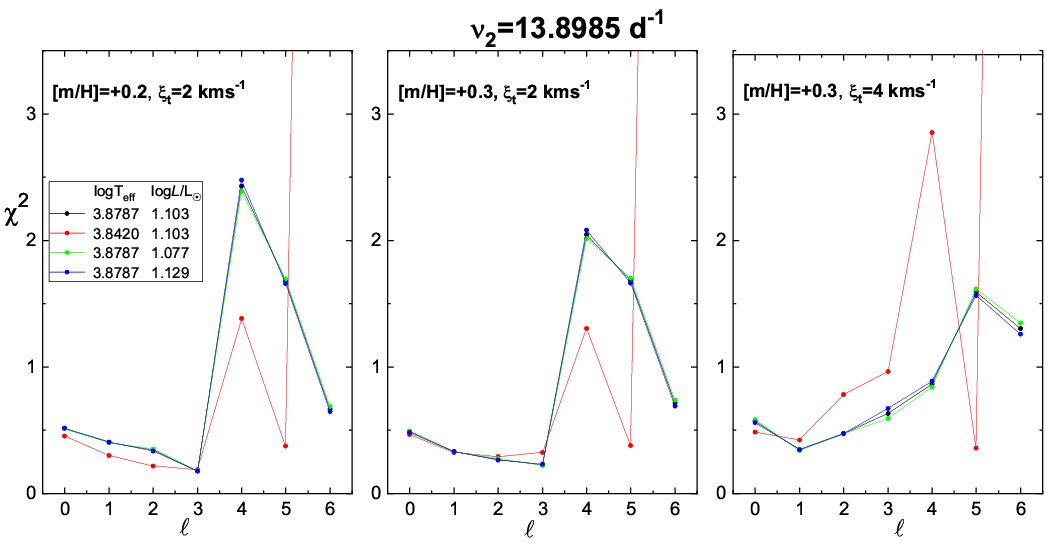}
	\caption{A similar figure to Fig.\,5 but for the frequencies $\nu_1=10.7377$\,d$^{-1}$ and $\nu_1=13.8985$\,d$^{-1}$ of RV Ari.}
	\label{fig6}
\end{figure*}

The period ratio corresponding to the frequencies $\nu_1$ and $\nu_2$ amounts to  0.77343 for AE UMa
and 0.77256 for RV Ari. These values strongly suggest that in each star $\nu_1$ corresponds to a radial fundamental mode
and $\nu_2$ to a radial first overtone. In this section, we independently verify this hypothesis using the method of mode 
identification based on the photometric amplitudes  and phases. To this end, we use  time-series photometry
in the Str\"omgren $uvby$ passbands made by \citet{Rodriguez1992}. In Table\,\ref{tab:amp_phas_Rod},
we give the amplitudes and phases derived from these data.  
\begin{table*}
	\centering
	\caption{The photometric amplitudes and phases in the $uvby$ passbands for the two dominant frequencies of AE UMa and RV Ari
	 determined from the data of \citet{Rodriguez1992}. The last column give the number of observational data points $N$.}
	\label{tab:amp_phas_Rod}
	\begin{tabular}{r r r l r l  r l r l c} 
		\hline
		star   &  frequency         &   $A_\mathrm{u}$ & $\phi_\mathrm{u}$&   $A_\mathrm{v}$ & $\phi_\mathrm{v}$&   $A_\mathrm{b}$ & $\phi_\mathrm{b}$&   $A_\mathrm{y}$ & $\phi_\mathrm{y}$ & $N$\\
		& $[\mathrm d^{-1}]$        &    [mag]           &   [rad] &    [mag]           &   [rad] &    [mag]           &   [rad] &    [mag]             &   [rad] & \\
		\hline
AE UMa  & $\nu_1$  & 0.2312(17)  & 4.301(7)  & 0.2941(16) & 4.186(5)  & 0.2569(14) &  4.191(5)  & 0.2112(15)  & 4.176(7) & 229\\
	            & $\nu_2$  & 0.0411(16)  & 4.892(40) & 0.0508(16) & 4.828(32) & 0.0429(14) &  4.841(33) & 0.0348(14)  & 4.861(42) & \\
		\hline
RV Ari  & $\nu_1$  & 0.2402(42)  & 2.181(19)  & 0.3083(39) & 2.103(14)  & 0.2657(34) &  2.093(14)  & 0.2213(33)  & 2.072(16) & 140\\
                & $\nu_2$  & 0.0730(43)  & 3.115(61) & 0.0909(40) & 3.018(45) & 0.0809(35) &  3.012(44) & 0.0613(34)  & 3.048(56) & \\
		\hline
	\end{tabular}
\end{table*}

Here, we apply the method of \citet{JDD2003} based on a simultaneous determination of the mode degree $\ell$, 
the intrinsic mode amplitude $\varepsilon$ multiplied by $Y^m_\ell(i,0)$ and the non-adiabatic parameter $f$ for 
a given observed frequency.
The numbers $\ell$ and $m$ are the spherical harmonic degree and the azimuthal order, respectively, and $i$  is the  inclination angle.

The intrinsic amplitude $\varepsilon$ of a mode is defined by the formula:
$$\delta r(R,\theta,\varphi)= R {\rm Re}\{ \varepsilon Y_\ell^m(\theta,\varphi)
{\rm e}^{-{\rm i}\omega t}\}, \eqno(2)$$
which gives the relative local radial displacement of the surface element caused by a pulsational mode
with the angular frequency $\omega$. Other symbols have their usual meanings.
The corresponding parameter $f$ is defined by  changes of the bolometric flux, ${\cal F}_{\rm bol}$ as
$$\frac{ \delta {\cal F}_{\rm bol} } { {\cal F}_{\rm bol} }= {\rm	Re}\{ \varepsilon f Y_\ell^m (\theta,\varphi) {\rm e} ^{-{\rm i} \omega t} \}.\eqno(3)$$
Both, $\varepsilon$ and $f$ have to be regarded as complex numbers because pulsations are non-adiabatic.
The theoretical values of $f=(f_R,~f_I)$ are derived from linear non-adiabatic computations of stellar pulsations 
whereas $\varepsilon$ is indeterminable under linear theory.  

In the linear and zero-rotation approximation, the theoretical expression for the complex amplitude
of the relative total flux variation in a passband $\lambda$, 
for a given pulsational mode, can be written in the form \citep[e.g.][]{JDD2003, JDD2005}:
$${\cal A}_{\lambda}= {\tilde\varepsilon} \left( {\cal D}_{\ell}^{\lambda}  f +{\cal E}_{\ell}^{\lambda} \right), \eqno(4)$$
where
$${\tilde\varepsilon}\equiv \varepsilon Y^m_{\ell}(i,0),\eqno(5a)$$
$${\cal D}_{\ell}^{\lambda} = b_{\ell}^{\lambda} \frac14
\frac{\partial \log ( {\cal F}_\lambda |b_{\ell}^{\lambda}| ) }
{\partial\log T_{\rm{eff}}}, \eqno(5b)$$
$${\cal E}_{\ell}^{\lambda}= b_{\ell}^{\lambda} \left[ (2+\ell
)(1-\ell ) -\left( \frac{\omega^2 R^3}{G M}
+ 2 \right) \frac{\partial \log ( {\cal F}_\lambda
	|b_{\ell}^{\lambda}| ) }{\partial\log g} \right],\eqno(5c)$$
and
$$b_{\ell}^{\lambda}=\int_0^1 h_\lambda(\mu) \mu P_{\ell}(\mu) d\mu.\eqno(5d)$$
The term ${\cal D}_{\ell}^\lambda$ describes the temperature effects and ${\cal E}_{\ell}^\lambda$ combines
the geometrical and pressure effect. $G,~M,~R$ have their usual meanings.
In equation Eq.\,5d, $h_\lambda(\mu)$ stands for the limb darkening law and $P_\lambda(\mu)$ is the Legendre polynomial.
The partial derivatives of ${\cal F}_\lambda(T_{\rm eff},\log g)$
in ${\cal D}_{\ell}^\lambda$ and ${\cal E}_{\ell}^\lambda$
as well as $b_{\ell}^{\lambda}(T_{\rm eff},\log g)$ and its derivatives have to be calculated from  model atmospheres.
Their values are sensitive to the metallicity [m/H] and microturbulent velocity $\xi_t$.
Here, we used Vienna model atmospheres \citep{Heiter2002} that include turbulent convection treatment from \citet{Canuto1996}.
For the limb darkening law, $h_\lambda(\mu)$, we computed coefficients assuming the non-linear, four-parametric formula 
of \citet{Claret2000}. The values of the photometric amplitudes  and phases themselves
are given by $A_{\lambda}=|{\cal A}_\lambda|$ and $\varphi_{\lambda}= arg({\cal A}_\lambda)$, respectively.

Next, the system of equations (4) for the four passband $uvby$
was solved for a given $\ell$ and $(T_{\rm eff},\log g)$ to determine $\tilde\varepsilon$ and $f$.
We considered the degree $\ell$ and associated complex values of $\tilde\varepsilon$ and $f$ as most probable if there is a clear
minimum in the difference between the calculated and observed photometric amplitudes and phases. 
The goodness of the fit  is measured by:
$$\chi^2=\frac1{2N-N_p} \sum_{i=1}^N  \frac{ \left|{\cal A}^{obs}_{\lambda_i} - {\cal A}^{cal}_{\lambda_i}\right|^2 }
{ |\sigma_{\lambda_i}|^2},\eqno(6)$$
where the superscripts $obs$ and $cal$  denote the observed and calculated complex amplitude 
${\cal A}_\lambda=A_\lambda {\rm e}^{{\rm i}\varphi_\lambda}$, respectively.
$N$ is the number of passbands $\lambda$ and $N_p$ is the number of parameters to be determined. 
$N_p=4$ because there are two complex parameters, $\tilde\varepsilon$ and $f$. 
The observational errors $\sigma_{\lambda}$ are computed as
$$|\sigma_\lambda|^2= \sigma^2 (A_{\lambda})  +  A_{\lambda}^2 \sigma^2(\varphi_\lambda), \eqno(7)$$
where $\sigma  (A_{\lambda})$ and $\sigma (\varphi)$  are the errors of the observed amplitude and phase 
in a passband $\lambda$, respectively.

In Fig.\,5, we show the values of the discriminant $\chi^2$ as a function of $\ell$ for the two frequencies of AE UMa; top panels are for $\nu_1$ and bottom panels for $\nu_2$. We considered several values of $T_{\rm eff}$ and $\log L/$L$_{\sun}$ within the observed error box.  We also checked the effect of the atmospheric metallicity [m/H] and microturbulent velocity $\xi_t$. 
In case of the dominant  mode,  for all values of ($\log T_{\rm eff},~\log L/{\rm L}_{\sun}$) and all considered pairs 
of  $\left({\rm [m/H]},~\xi_t\right)$ the clear minimum of $\chi^2$ is at $\ell=0$. Thus, there is no doubt that $\nu_1$ is a radial mode.
For the second frequency, the minimum at $\ell=0$ is not significantly smaller than at the other $\ell$'s. However, 
given that the visibility of pulsational modes decreases very rapidly with increasing $\ell$, it is reasonable to assume 
the $\ell=0$ identification for $\nu_2$.

The identification of $\ell$ of the two modes of RV Ari is presented in Fig.\,6.  Our photometric method clearly indicates
that the dominant mode is radial. For the second frequency the identification is not unambiguous. This is mostly because 
of much larger errors in the photometric amplitudes and phases of RV Ari (about three time larger  comparing to AE UMa).
It results from lower number  of observational data points and the fact that RV Ari is fainter than AE UMa. 
However, from the plot $\chi^2$ vs. $\ell$ for $\nu_2$, one can conclude that $\ell=0,1,2,3$ are equally possible. 
Combining this fact with the  period ratio and the largest visibility factor $b_{\ell}$ (cf. Eq. 5d) for $\ell=0$, 
it is safe to  assume for further analysis that  $\nu_2$ is also a radial mode. 

\section{Complex seismic modelling}
The complex seismic modelling consists in the simultaneous matching of pulsational frequencies and the corresponding
values of the non-adiabatic parameter $f$. 
The parameter $f$ gives the relative amplitude of the radiative flux perturbation at the photosphereic level.
Its theoretical value for a given pulsational mode is obtained from non-adiabatic computations of stellar pulsations
and it is complex because there is a phase shift between the radiative flux variation  and radius variation.
In the case of $\delta$ Sct stellar models, the theoretical values of $f$ are very sensitive  
to the efficiency of convection in the outer layers and to opacity data \citep[e.g.,][]{JDD2003,JDD2023}.  
By comparing the theoretical and empirical values of $f$, one can get valuable constraints on the physical conditions 
inside the star.  Thus, the parameter $f$ is a seismic tool that carries information about the stellar interior and that is independent and complementary to the pulsation frequency.

The empirical values of $f$ and $\tilde\varepsilon$ were determined from the amplitudes and phases in the $uvby$ passbands
using the method outlined in Sect.\,4. 
In the case of radial modes $Y_{\ell}^m(i,0)=1$ and we have the value of $\varepsilon$  itself (cf. Eq.\,5a), i.e., we can say what are 
the percentage changes in the radius caused by each pulsation mode (cf. Eq.\,2).
As before, we adopted Vienna model atmospheres \citep{Heiter2002}.  
Models with the microturbulent velocity $\xi_t = 4~\kms$ gave the smallest errors in the empirical values of $\varepsilon$ and $f$.
Therefore, we adopted $\xi_t = 4~\kms$ whereas the atmospheric metallicity [m/H] was changed consistently 
with the metallicity $Z$ in evolutionary computations.
\begin{table*}
	\centering
	\setlength{\tabcolsep}{2pt}
	\caption{The expected (E$X$) and median (Med) values of the parameters of the {\bf HSB} seismic models of the stars AE UMa 
		and RV Ari. The parameters were obtained  from the Bayesian analysis based on Monte-Carlo simulations assuming 
		the OPAL opacities and the Vienna model atmospheres with the  microturbulent velocity of $\xi_t$=4 kms$^{-1}$.}
	\label{tab:expected_media_HSB}
	\begin{tabular}{ccccccccccccccc}
		\hline
		star & value & $M$ & $Z$ & $X_0$ & $\alpha_{\rm{mlt}}$ & age & $V_{\rm{0,rot}}$ & $V_{\rm{rot}}$ &
		$\log (T_{\rm eff}/{\rm K})$ & $\log{L/{\rm L}_{\sun}}$ & $\alpha_{\rm{ov}}$\\
		&    & [M$_{\sun}$] &     &       &        &  [Gyr]      & [km$\cdot$s$^{-1}$] & [km$\cdot$s$^{-1}$] &
		&                          &                   \\
		\hline	
		AE UMa & E$X$ & 1.567(41)  & 0.0130(10)  & 0.698(17)  & 0.43(17)  & 1.587(73) &
		18.4(11.5)  & 18.4(11.5) & 3.8613(36)  & 1.091(20)  & 0.024(18)   \\
		&&&&&&&&&\\
		& Med &$1.564^{+0.049}_{-0.040}$ & $0.0131^{+0.0010}_{-0.0009}$ &   
		$0.697^{+0.023}_{-0.018}$ &    $0.44^{+0.15}_{-0.18}$ &   $1.574^{+0.078}_{-0.057}$ &
		$17.9^{+12.9}_{-12.5}$ &    $17.8^{+12.9}_{-12.5}$ &
		$3.8617^{+0.0031}_{-0.0040}$ &    $1.092^{+0.020}_{-0.022}$ &   
		$0.021^{+0.020}_{-0.012}$     \\
		&&&&&&&&&\\
		\hline
		RV Ari & E$X$ & 1.629(43)  & 0.0178(18)  & 0.689(20)  & 0.53(7)  &  1.565(72)  &
		23.9(15.5) & 23.4(15.2)  & 3.8484(50)  & 1.093(27)  &  0.004(2)   \\
		&&&&&&&&&\\
		& Med & $1.631^{+0.037}_{-0.044}$ & $0.0179^{+0.0020}_{-0.0021}$ &   
		$0.685^{+0.024}_{-0.015}$ &    $0.53^{+0.07}_{-0.08}$ &  $1.552^{+0.085}_{-0.059}$ &
		$22.1^{+20.8}_{-15.1}$ & $21.5^{+20.6}_{-14.7}$ & $3.8501^{+0.0027}_{-0.0075}$ &   
		$1.100^{+0.017}_{-0.038}$ &    $0.004^{+0.002}_{-0.002}$ \\
		&&&&&&&&&\\
		\hline
	\end{tabular}\\
\end{table*}
\begin{table*}
	\centering
	\setlength{\tabcolsep}{2pt}
	\caption{The same as in Table\,5 but for the {\bf OC} seismic models of the stars AE UMa and RV Ari.}
	\label{tab:expected_media_OC}
	\begin{tabular}{ccccccccccccccc}
		\hline
		star & value & $M$ & $Z$ & $X_0$ & $\alpha_{\rm{mlt}}$ & age & $V_{\rm{0,rot}}$ & $V_{\rm{rot}}$ &
		$\log (T_{\rm eff}/{\rm K})$ & $\log{L/{\rm L}_{\sun}}$ & $\alpha_{\rm{ov}}$\\
		&    & [M$_{\sun}$] &     &       &        &  [Gyr]      & [km$\cdot$s$^{-1}$] & [km$\cdot$s$^{-1}$] &
		&                          &                   \\
		\hline	
		AE UMa & E$X$ & 1.532(52)  & 0.0120(11)  & 0.701(20)  & 0.33(17)  & 1.706(114) &
		17.7(13.3)  & 17.5(13.2) & 3.8578(69)  & 1.070(31)  & 0.060(36)   \\
		&&&&&&&&&\\
		& Med &$1.535^{+0.050}_{-0.058}$ & $0.0119^{+0.0014}_{-0.0011}$ &   
		$0.698^{+0.026}_{-0.017}$ &    $0.33^{+0.16}_{-0.17}$ &   $1.683^{+0.158}_{-0.094}$ &
		$14.6^{+18.0}_{-10.6}$ &    $14.3^{+18.0}_{-10.5}$ &
		$3.8587^{+0.0055}_{-0.0076}$ &    $1.072^{+0.027}_{-0.032}$ &   
		$0.055^{+0.049}_{-0.032}$     \\
		&&&&&&&&&\\
		\hline
		RV Ari & E$X$ & 1.657(26)  & 0.0180(11)  & 0.692(12)  & 0.59(7)  &  1.528(49)  &
18.6(15.4) & 18.1(15.0)  & 3.8525(49)  & 1.115(24)  &  0.004(3)   \\
&&&&&&&&&\\
& Med & $1.657^{+0.027}_{-0.026}$ & $0.0180^{+0.0012}_{-0.0011}$ &   
$0.693^{+0.012}_{-0.015}$ &    $0.59^{+0.07}_{-0.07}$ &  $1.520^{+0.054}_{-0.041}$ &
$16.5^{+18.7}_{-12.4}$ & $15.9^{+18.3}_{-11.9}$ & $3.8532^{+0.0043}_{-0.0041}$ &   
$1.118^{+0.021}_{-0.027}$ &    $0.004^{+0.003}_{-0.003}$ \\
&&&&&&&&&\\
		\hline
	\end{tabular}\\
\end{table*}

We performed an extensive complex seismic modelling of AE UMA and RV Ari by fitting the two radial mode frequencies and the non-adiabatic parameter $f$ for the dominant modes.  In the case of both stars, the second modes  had too low amplitudes 
to determine  the empirical values of $f$ with enough accuracy. To find seismic models, we used the Bayesian analysis based 
on Monte Carlo simulations. Our approach is shortly described in Appendix\,B and the details can be found in \citet{JDD2022,JDD2023}.
Here, we just recall the adjustable parameters: mass $M$, initial hydrogen abundance $X_0$, metallicity $Z$, 
initial rotational velocity $V_{\rm rot,0}$, convective overshooting parameter $\alpha_{\rm ov}$ and the mixing length 
parameter $\alpha_{\rm MLT}$. Because only computations with the OPAL data give consistent results with the observational values 
of $(T_{\rm eff},~L/{\rm L}_{\sun})$ \citep{JDD2023}, we adopted these opacities in all computations.
At lower temperature range, i.e., for $\log T<3.95$,  opacity data from \citet{Ferguson2005} were used.

Evolutionary computations were performed using  the Warsaw-New Jersey code, \citep[e.g.,][]{Pamyatnykh1999}.
The code takes into account the mean effect of the centrifugal force, assuming solid-body rotation
and constant global angular momentum during evolution. Because both stars are slow rotators, neglecting 
differential rotation is justified. 
Convection in stellar envelope is treated in the framework of standard mixing-length theory (MLT) and its efficiency
is measured by the mixing length parameter $\alpha_{\rm MLT}$. The solar chemical mixture was adopted 
from \citet{Asplund2009} and the OPAL2005 equation of state was used \citep{Rogers1996,Rogers2002}.
Overshooting from a convective core in the code is implemented according to \citet{DziemPamy2008}. 
Their prescription takes into account both the distance of overshooting $d_{\rm ov}=\alpha_{\rm ov} H_p$,
where $H_p$ is the pressure scale height and $\alpha_{\rm ov}$ is a free parameter, 
as well as a hydrogen profile $X(m)$ in the overshoot layer.

Non-adiabatic stellar pulsations were computed using a linear code of \citet{Dziembowski1977a}. The code assumes
that  the convective flux does not change during pulsations which is justified  if convection is not very efficient in the envelope.
The effects of rotation on pulsational frequencies are taken into account up to the second order in the framework of perturbation theory.

We calculated evolutionary and pulsational models  for each set of randomly selected parameters
$\left(M, ~X_0, ~Z, ~V_{\rm rot,0}, ~\alpha_{\rm MLT}, ~\alpha_{\rm ov}\right)$. 
The number of simulations  was about 360\,000 for each  star.  In the case of initial hydrogen abundance $X_0$, 
we assumed a beta function $B(2, 2)$ as a prior probability to limit its value to the reasonable range, i.e., from 0.65 to 0.75 
with $X_0 = 0.7$ as the most probable. For other parameters we used uninformative priors, i.e., uniform distributions.
The vast majority of  our seismic models of the two stars, that have the values of ($T_{\rm eff},~L/{\rm L}_{\sun}$)  consistent
with the observational determinations, are already at the beginning  of hydrogen-shell burning (HSB).
Only a small fraction of seismic models with proper values of  ($T_{\rm eff},~L/{\rm L}_{\sun}$)  is  an overall contraction (OC) phase,
at its very end. 
In all seismic modes both radial modes, fundamental and first overtone, are unstable in both stars.
In Table\,\ref{tab:expected_media_HSB}, we give the expected and median values of determined parameters of the seismic models 
in the HSB phase for the two HADS stars. 
The errors in parentheses at the expected values are standard deviations. The median errors were estimated from 
the 0.84 and 16 quantiles, which correspond to the one standard deviation from the mean value in the case of a normal distribution.
Table\,\ref{tab:expected_media_OC} contains the same statistics for the seismic  models
in the OC phase.  The corresponding corner plots and histograms for the HSB seismic models are presented in Appendix B, in Figs.\,B1-B4.
The histograms for the OC seismic models look qualitatively similar.
\begin{figure*}
	\centering
	\includegraphics[clip,width=0.49\linewidth]{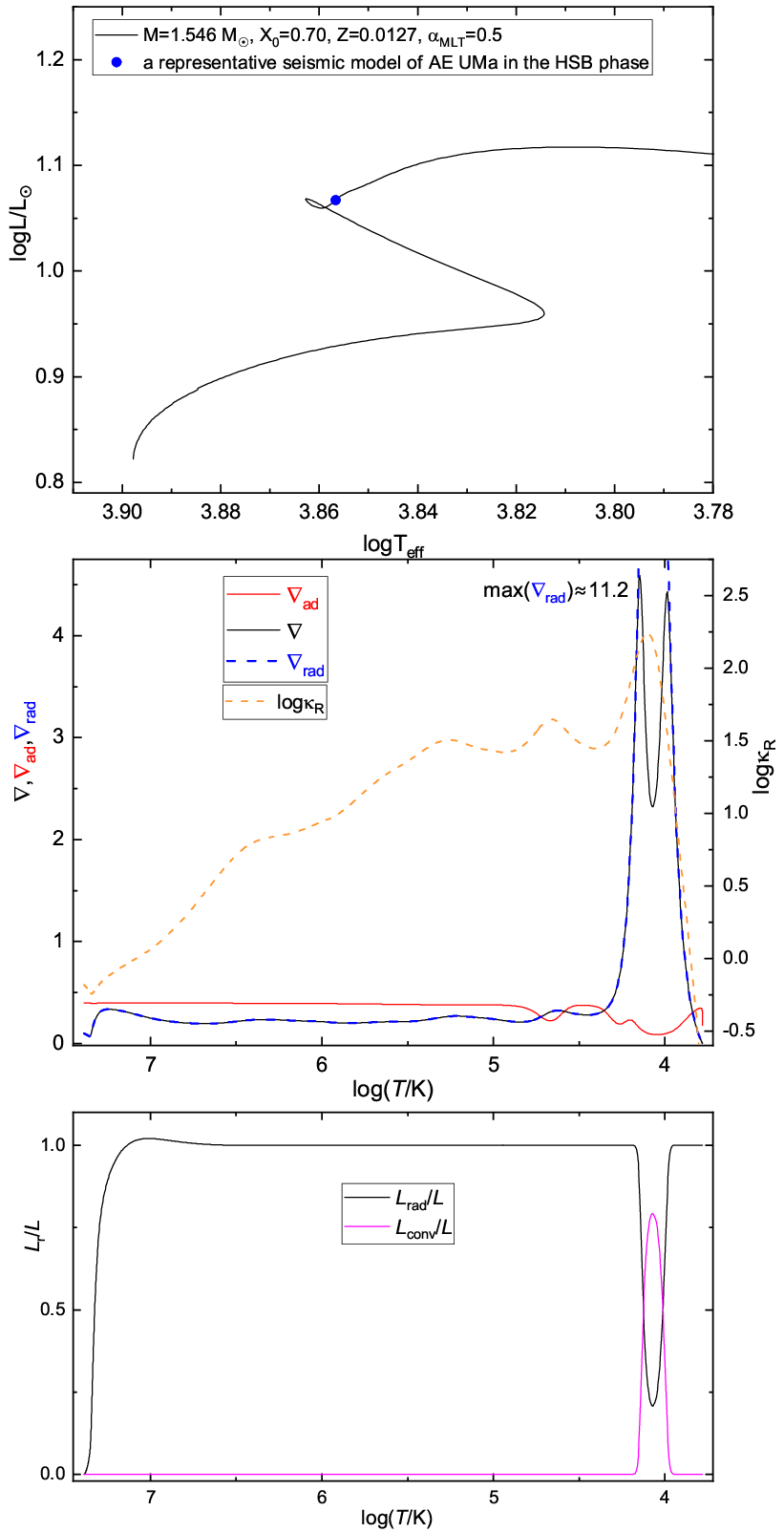}
	\includegraphics[clip,width=0.49\linewidth]{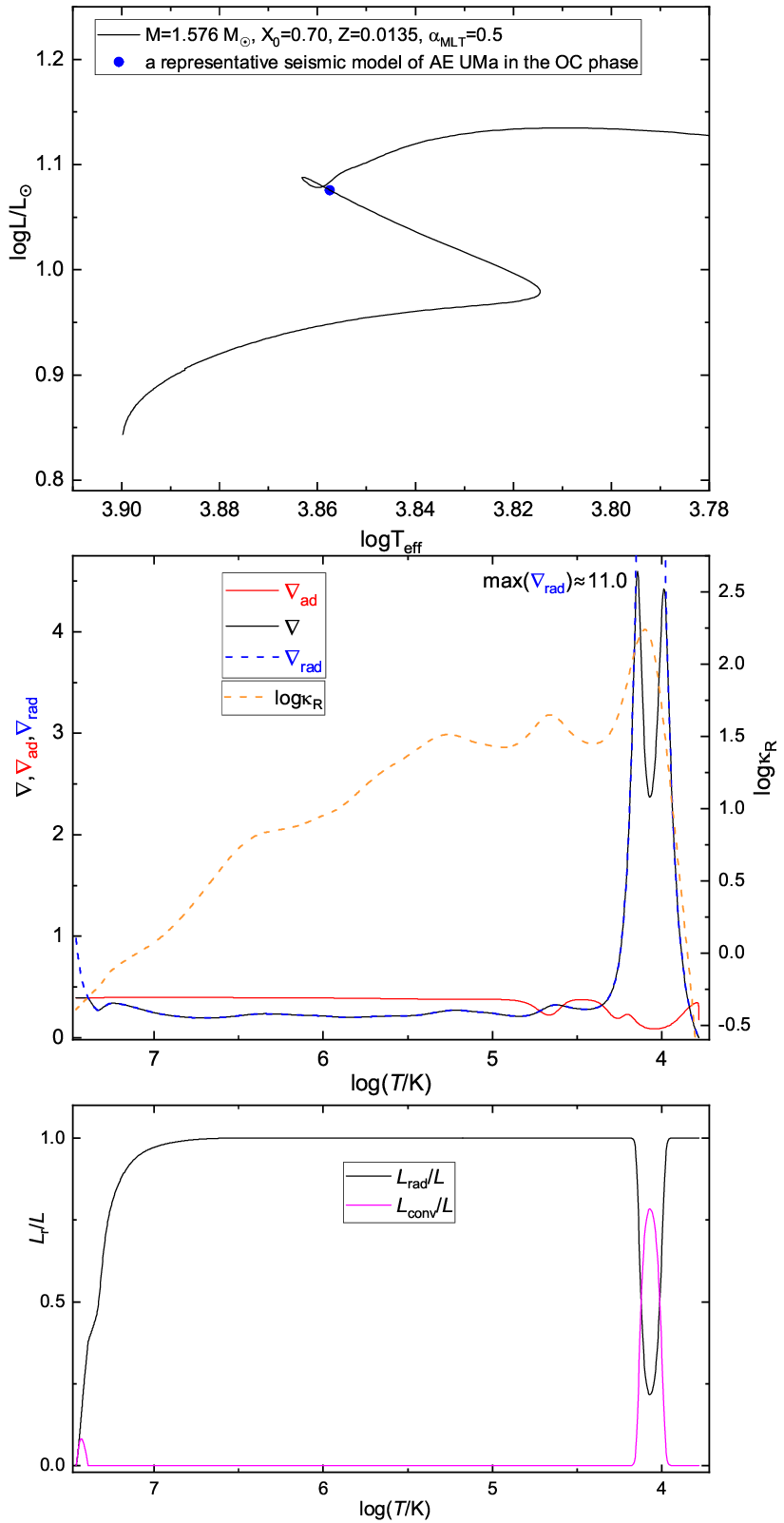}
	\caption{The structure of two seismic models of AE UMa. Their positions on the HR diagram is indicated in top panels.
		The models reproduce two radial mode frequencies and the non-adiabatic parameter $f$ for the dominant mode.
		Left panel presents the model in the HSB phase and right panel the model in the overall contraction.
		Middle panels show the run of main three gradients and the mean Rosseland opacity inside the models whereas
		bottom panels show the local radiative and convective luminosity. }
	\label{fig:seismic_structure}
\end{figure*}
Two stars have a very similar position in the HR diagram but RV Ari is more massive and has higher metallicity than AE UMa.
The age of both HADS stars is quite similar and amounts to about 1.6\,Gyr, if the stars are in the HSB phase of evolution. 
Seismic models in the OC phase are about 100\,Myr  older in the case of AE UMa and about 30\,Myr younger in the case of RV Ari.

We obtained, that convection in the outer layers of the two stars is not very efficient. For the HSB seismic models, the mixing length parameter amounts to about 0.4 for AE UMa and about 0.5 for RV Ari. The OC seismic models
of AE UMa and RV Ari, have $\alpha_{\rm MLT}$ of  about 0.3 and 0.6, respectively. 

The most striking result is a very small overshooting from the convective core. 
This result may indicate that the overshooting parameter should depend on time (evolution) and, presumably, 
should scale with a mass and size of the convective core.

In Fig.\,7, we show the structure of two representative complex seismic models of AE UMa. Seismic models of RV Ari have 
qualitatively the same structure.  Left panels show the model in the HSB phase  and right panels  the model in the OC phase. 
Both models have the following common parameters: $X_0=0.70,~\alpha_{\rm ov}=0.0$ and $\alpha_{\rm MLT}=0.5$.
The other parameters are: $M=1.546M_{\sun},~Z=0.0127,~\log T_{\rm eff}=3.8566,~\log L/{\rm L}_{\sun}=1.067$
for the HSB model and $M=1.576M_{\sun},~Z=0.0135,~\log T_{\rm eff}=3.8575,~\log L/{\rm L}_{\sun}=1.076$ for the OC model.
In the top panels, we show the positions of these models on the HR diagram with the corresponding evolutionary tracks.
Middle panels show the run of main gradients (actual $\nabla$, radiative $\nabla_{\rm rad}$, adiabatic $\nabla_{\rm ad}$) 
and the mean Rosseland opacity $\kappa$ inside each model. The local radiative and convective luminosity is presented 
in the bottom panels.  

In the case of HSB model, the small helium core  has the size of 3\% of  the stellar radius  and the 
interior up to  $\log (T/{\rm K}) \approx 4.8$ is radiative. The whole energy comes from hydrogen-burning in the shell proceeding 
via the CNO cycle. An overproduction of energy in the shell $L_r/L>1$ is used for expansion of the envelope.
The radiative gradient become very large around $\log (T/{\rm K}) =4.0$ where the actual gradient split into the two maxima 
corresponding to the hydrogen ionization and first ionization of helium. In this narrow layer the local convective luminosity
becomes important. In the zone of  second helium ionization  $\log (T/{\rm K}) \approx 4.7$, where pulsational driving occurs,
 $\nabla$ is only slightly larger than $\nabla_{\rm ad}$
and the local convective luminosity is zero. In the case of the OC models, the hydrogen is burned in a small convective core  
with the radius of about $3\% R$. The structure of the OC model above the core is very similar to that of the HSB model.
\begin{table}
	\centering
	\caption{The modal values of the empirical intrinsic mode amplitude $|\varepsilon|$ (a fraction of the radius changes)
		and the resulting radial velocity amplitude due to pulsations for the two radial modes of AE UMa and RV Ari.}
	\label{tab:eps_model}
	\begin{tabular}{c c l c l l} 
		\hline
		&   \multicolumn{2}{c}{$\nu_1$} & \multicolumn{2}{c}{$\nu_2$} \\
		&   \multicolumn{2}{c}{$\ell=0,~p_1$} & \multicolumn{2}{c}{$\ell=0,~p_2$} \\
		\hline
		star   &  $|\varepsilon|$   &   $A(V_{\rm puls})$ & $|\varepsilon|$   &  $A(V_{\rm puls})$  \\
		&             &     [km$\cdot$s$^{-1}$]  &  & [km$\cdot$s$^{-1}$] \\
		\hline
		AE UMa  &   0.0189(24)  &  17.7(2.2) &  0.0017(6)  &  2.0(7)   \\
		&&&&\\
		\hline
		RV Ari  &   0.0153(4)   &  13.9(4)   &  0.0029(20) &  3.4(2.3)  \\	
		&&&&\\
		\hline
	\end{tabular}
\end{table}

As we mentioned at the beginning of this section, from our analysis, we obtained also the empirical values of the intrinsic mode amplitude $\varepsilon$ for both 
radial modes. These values  cannot be compared  with theoretical predictions, because we use the linear theory,  but it provides 
us with an estimate of the relative radius changes and the expected amplitude of radial velocity variations  $A(V_{\rm puls})$.
In Table\,\ref{tab:eps_model}, we give the values of $\varepsilon$ and the corresponding radial velocity amplitude for the two 
radial modes of both stars. These are the modal values, i.e., is the most frequently occurring  values in computed  seismic models. 
As one can see, the radial fundamental modes of AE UMa and RV Ari cause the radius changes of about 1.9\% and 1.5\%, respectively.
In turn, these radius changes cause the radial velocity variations with an amplitude of about 18 and 14\,$\kms$, respectively.
The first overtone modes have much smaller radius variations of about 0.2\% and 0.3\% for AE UMa and RV Ari, respectively.
The amplitude of radial velocity variations caused by the first overtone modes are about 2.0 and 3.4\,$\kms$ for AE UMa and RV Ari, respectively.

\section{Including the nonradial mode of RV Ari into seismic modelling}
\begin{table*}
	\centering
	\caption{Examples of seismic models of RV Ari with the characteristic of nonradial modes $\ell=1, 2$ having frequencies
	 closest to the observed value $\nu_3=13.6118$\,d$^{-1}$. All seismic models have $\alpha_{\rm ov}=0.0$
	 and were computed with the OPAL opacities. The 8th column indicates the evolutionary phases. The last three columns contain a ratio of  the energy in gravity propagation zone to the total kinetic energy, the instability parameter $\eta$ and the Ledoux constant.
	 The models indicated with asterisks will be discussed later in the text. }
	 	\label{tab:nonrad_modes}
	\begin{tabular}{cccccccccccccccccc}
		\hline
    $M$ & $Z$ & $X_0$  &   $\alpha_{\rm MLT}$  & $\log (T_{\rm eff}/{\rm K})$ & $\log{L/{\rm L}_{\sun}}$  & age & phase &
     $\nu_{\rm rot}$  & $\nu_{\rm puls}$ & mode & $E_{\rm k,g}/E_{\rm k}$ & $\eta$ & $C_{\rm n\ell}$  \\
      $[{\rm M}_{\sun}]$ &  &   &  & &  & [Gyr]  &  & [d$^{-1}$] & [d$^{-1}$]  &  & & & \\
		\hline	
 1.590 & 0.015 & 0.70 & 0.55 & 3.8469 & 1.081 & 1.6435 & HSB & 0.206 & 13.69773 & $\ell=1,g_{7}$   & 0.82 & 0.090 & 0.426 \\
       &       &      &      &        &       &        &     &       & 13.87880 & $\ell=2,g_{12}$  & 0.77 & 0.089 & 0.130 \\
       \hline
1.624{\bf *} & 0.016 & 0.70 & 0.55 & 3.8485 & 1.093 & 1.5973 & HSB & 0.203 & 13.70427 & $\ell=1,g_{6}$   & 0.83 & 0.089 & 0.432 \\
       &       &      &      &        &       &        &     &       & 13.44373 & $\ell=2,g_{12}$  & 0.76 & 0.090 & 0.147 \\
       \hline
 1.652 & 0.017 & 0.70 & 0.55 & 3.8491 & 1.100 & 1.5684 & HSB & 0.202 & 14.16947 & $\ell=1,g_{5}$   & 0.43 & 0.088 & 0.216 \\
       &       &      &      &        &       &        &     &       & 13.35265 & $\ell=2,g_{11}$  & 0.76 & 0.090 & 0.151 \\
       \hline
 1.640 & 0.019 & 0.68 & 0.64 & 3.8493 & 1.098 & 1.5205 & HSB & 0.210 & 13.37437 & $\ell=1,g_{5}$   & 0.90 & 0.089 & 0.468 \\
       &       &      &      &        &       &        &     &       & 13.34091 & $\ell=2,g_{10}$  & 0.75 & 0.089 & 0.152 \\
       \hline
1.557{\bf *} & 0.018 & 0.68 & 0.50 & 3.8274 & 0.995 & 1.7090 &  OC & 0.399 & 14.3346 & $\ell=1,g_{1}$   & 0.07 & 0.066 & 0.030 \\
       &       &      &      &        &       &        &     &       & 13.2824 & $\ell=2,g_{4}$   & 0.61 & 0.076 & 0.144 \\
		\hline
	\end{tabular}\\
\end{table*}

\begin{figure*}
	\centering
	\includegraphics[clip,width=0.49\linewidth,height=15.5cm]{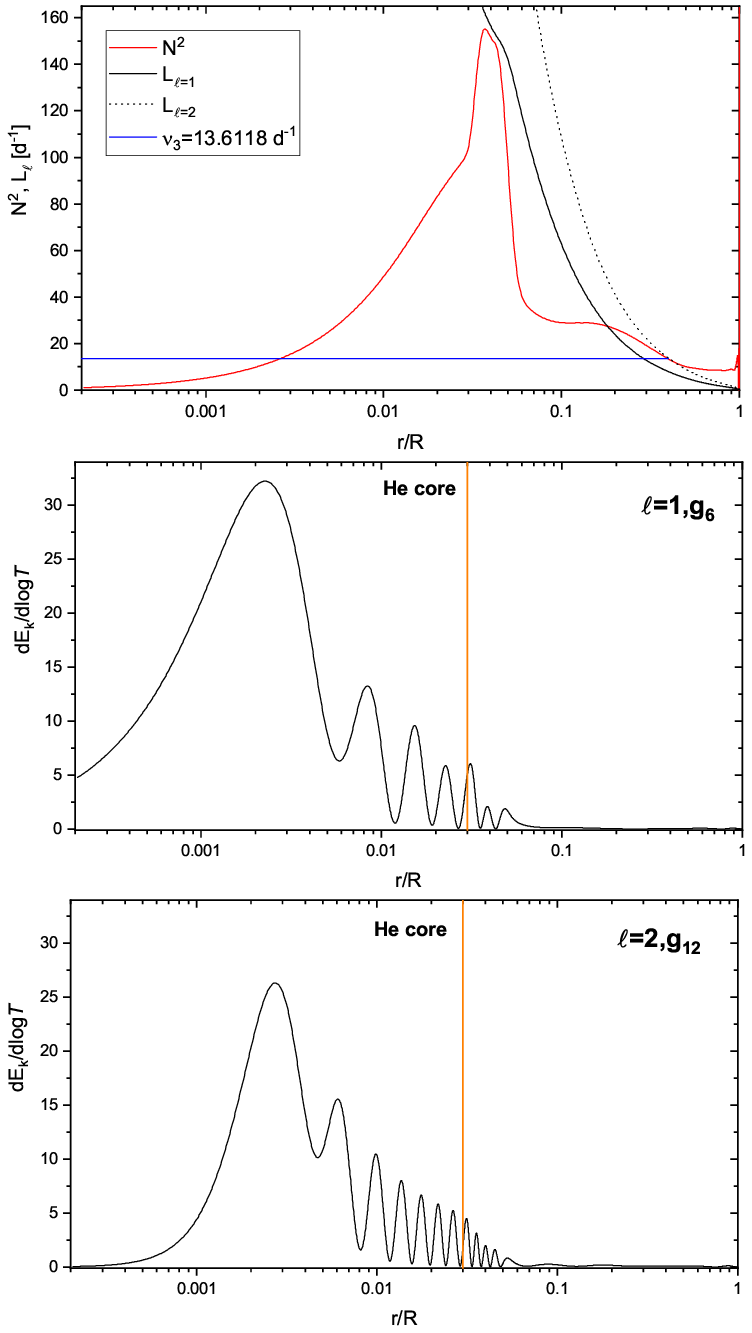}
	\includegraphics[clip,width=0.49\linewidth,height=15.5cm]{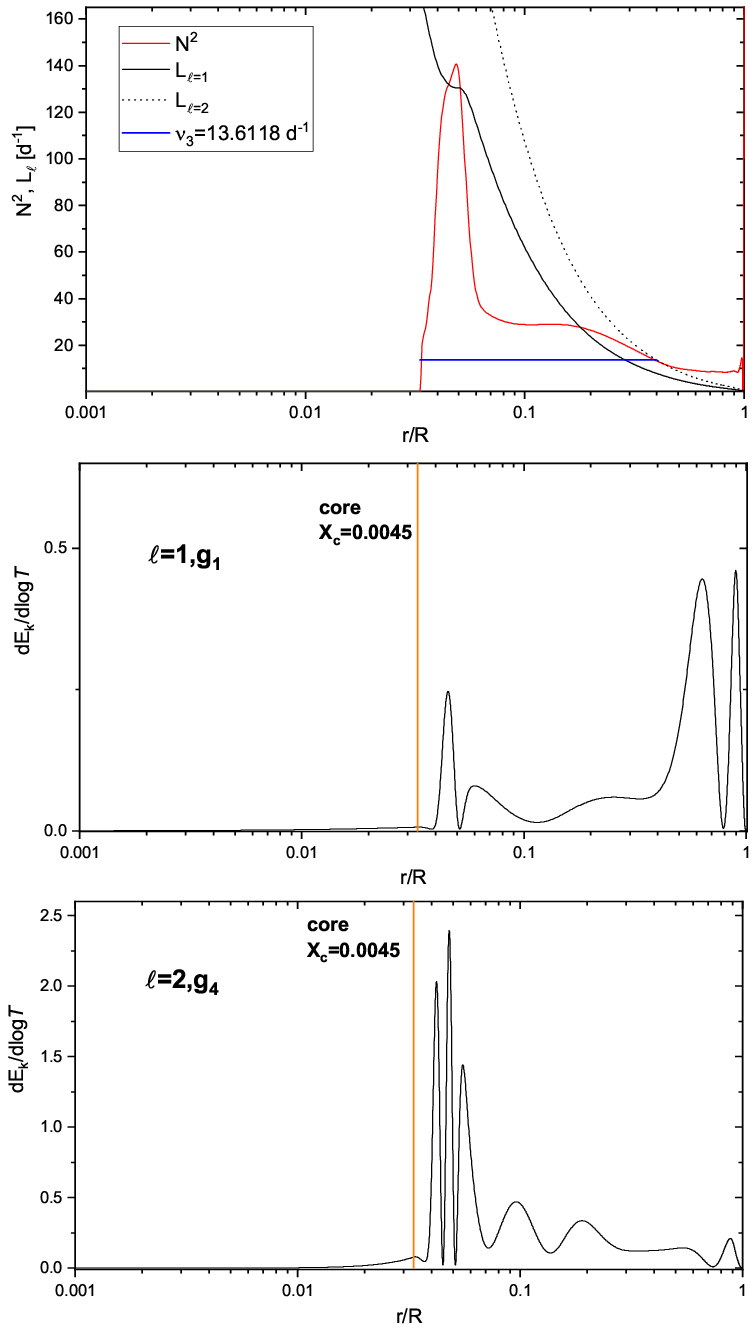}
	\caption{Top panels: the propagation diagram for the HSB (left) and OC (right) seismic models of RV Ari with the parameters
		given in the 2nd and 5th line of Table\,\ref{tab:nonrad_modes}. The horizontal line indicates the value of the observed frequency $\nu_3=13.61183$\,d$^{-1}$. Lower panels: the distribution of the kinetic energy density for nonradial modes listed in 
		Table\,\ref{tab:nonrad_modes}. The vertical lines mark the edge of a core.}
	\label{fig:seismic_nonrad_modes}
\end{figure*}

Our results of the Fourier analysis of the space TESS data confirmed that the third frequency of RV Ari,
$\nu_3=13.61183$\,d$^{-1}$, proposed by \citet{Pocs2002},  is a real and independent signal. 
This frequency  can only be associated with a nonradial mode because of its proximity
to the second frequency $\nu_2=13.899137$\,d$^{-1}$ which corresponds to the first overtone radial mode.
The $\nu_3$ does not appear in the \citet{Rodriguez1992} data, so we cannot even try to identify 
its pulsational mode from the photometric amplitudes and phases. On the other hand, taking into account the very small amplitude of this frequency (more than 3 times smaller than the amplitude of $\nu_2$), we doubt that such an attempt would be successful. For this purpose, new time-series multi-colour photometry or spectroscopy is required.
However, guided by the fact that the visibility of modes in photometry decreases rapidly with the degree $\ell$,
$\nu_3$ is quite likely a dipole or quadrupole mode. Therefore, we will make such working hypothesis in this Section. 
 
In Table\,\ref{tab:nonrad_modes}, we list the parameters of the four HSB and one OC  seismic models of RV Ari 
with the main characteristic  of dipole and quadrupole modes having frequencies closest to the observed value 
of $\nu_3=13.61183$d$^{-1}$. 
We provide the ratio of the kinetic energy in the gravity propagation zone to the total kinetic energy $E_{\rm k,g}/E_{\rm k}$,
the normalized instability parameter $\eta$ and the Ledoux constant $C_{\rm n\ell}$. As one can see, despite of quite high 
frequencies, in the case of the HSB seismic models, the modes have a very strong gravity character with  $E_{\rm k,g}$ 
greater than 70\% of the total $E_{\rm k}$ in all cases but one. Only in the case of HSB model with $M=1.652$\,M$_{\sun}$, 
the mode $\ell=1,~g_5$   has this ratio slightly  below 0.5. In the case of the OC seismic model,  $\ell=1,~g_1$  is almost a pure 
pressure mode but its frequency is quite far from $\nu_3$.
The mode $\ell=2,~g_4$  is  mixed with  $E_{\rm k,g}$ of about 60\%. All modes are pulsational unstable.

In Fig.\,\ref{fig:seismic_nonrad_modes}, we show the propagation diagram (top panels) and the distribution of kinetic energy density 
of the dipole  and quadruple mode (lower panels) for the seismic models indicated with asterisk in Table\,\ref{tab:nonrad_modes}
(the 2nd and 5th model) .  
All quantities  are plotted as a function of the fractional radius in a logarithmic scale. The 2nd model is in the phase of hydrogen-shell
burning  whereas the 5th model is in the overall contraction phases.  The Lamb frequency $L_{\ell}$ was depicted for $\ell=1$ and 2. 
The horizontal line in the top panels corresponds to the observed frequency $\nu_3=13.61183$\,d$^{-1}$. In the case of the HSB model,
the maximum of the Brunt-V\"ais\"al\"a frequency $N^2$ occurs at the edge of a small helium core. The small convective core 
of the OC model is precisely defined by $N^2<0$. The core edges are marked as a vertical line.  The middle and bottom panels show 
the kinetic energy density  for the modes  $\ell=1$ and $\ell=2$, respectively. 
As one can see, the kinetic energy density of both modes of the HSB models is large and strongly concentrated within the helium core 
of a size of $0.03R$.  Modes with such property have a very strong potential for probing near-core conditions
and chemical composition. In the case of the OC model, the mode  $\ell=1, g_1$ is almost pure pressure and its kinetic energy 
concentrates in the outer layers. The mixed mode $\ell=2, g_4$  has the kinetic energy concentrated within the chemical gradient zone.

In the next step, we constructed seismic models that fit simultaneously the two radial mode frequencies, the complex parameter $f$
of the dominant mode and $\nu_3$ as a dipole axisymmetric mode.  Again, the Bayesian analysis based on Monte Carlo simulations 
was applied. 
We obtained very narrow ranges of determined parameters $M$,  $X_0$, $Z$ with errors about 2 to 5 times smaller that in Sect.\,5.
whereas the values of  overshooting and mixing length parameters are again around 0.0 and 0.5, respectively, with a similar uncertainty.
Interestingly, in the case of HSB seismic models our simulations converge to the two solutions for $(M,~X_0,~Z)$
with the following expected values:

~I$.~~~~M=1.644(7)\,{\rm M}_{\sun}, ~X_0=0.689(10),~Z=0.0186(4)$

II$.~~~~M=1.626(8)\,{\rm M}_{\sun}, ~X_0=0.696(7),~Z=0.0164(3)$

This dichotomy is most evident for metallicity $Z$. In the Appendix B, in Fig.\,B5, we plot the metallicity $Z$ 
as a function of the model number. As one can see, independently of the starting value, the simulations converge 
only to the two values of  $Z$ given above. 
In the first solution a dipole mode is always $g_7$ and in the second solution it is always $g_8$.

For OC seismic models we got one solution with a higher $Z$:

~I$.~~~~M=1.640(10)\,{\rm M}_{\sun}, ~X_0=0.693(7),~Z=0.0181(5)$,

\noindent and  a dipole mode is $g_4$.  As in the case of fitting the two radial modes, the OC seismic models that reproduce
also the third frequency $\nu_3$ as a dipole mode are a definite minority.

The interesting result is that the obtained range of the rotational velocity differs significantly between the HSB and OC seismic models.
For the HSB models, the expected value of the current rotation velocity is in the range $V_{\rm rot}\in (5,~46)\,\kms$ whereas 
for the OC models we obtained the range  $V_{\rm rot}\in (32,~47)\,\kms$. 
Thus, if  we had independent information on the rotation rate, e.g., from the rotational splitting of dipole modes, 
then perhaps a choice between HSB and OC siesmic models would be possible.

\section{Summary}

We presented the analysis of TESS space data and detailed complex  
seismic modelling  of the two high-amplitude $\delta$ Scuti stars AE UMa and RV Ari.
The Fourier analysis of the TESS light curves revealed the two well-known frequencies for each HADS. The important result 
is a confirmation of  the third frequency of RV Ari that has been detected from the ground-based photometry.  

The ratio of two dominant frequencies of AE UMa and RV Ari strongly indicates that two radial modes, fundamental
and first overtone, are excited in both stars.
We verified this hypothesis using the method of  mode identification based on the multi-colour photometric amplitudes and phases.
 
Our seismic modelling of the two HADS stars consisted in simultaneous fitting of the two radial mode frequencies as well as
the complex amplitude of relative bolometric flux variations of the dominant mode, the so-called parameter $f$.
To this end, the Bayesian analysis based on Monte-Carlo simulations was used. Our extensive  seismic modelling
allowed to constrain the global parameters as well as  free parameters.
The mixing length parameter $\alpha_{\rm MLT}$, that describes the efficiency of envelope convection, amounts
to about 0.3$-$0.6. Determination of the narrow range of $\alpha_{\rm MLT}$ was possible 
only due to the inclusion of the parameter $f$ into seismic modelling.
All models are in the post-main sequence phase of evolution, however  the question whether it is the HSB or OC phase 
cannot be unequivocally resolved. On the other hand, the HSB seismic models account for the vast majority, so it can be 
assumed that this phase is much more likely.

An interesting result is a very small value of the overshooting parameter $\alpha_{\rm ov}$, which describes the amount 
of mixing at the edge of convective core during the phase of main-sequence and overall contraction.  This result may be due 
to the fact that $\alpha_{\rm ov}$  is assumed to be constant during evolution. Presumably, it should depend on time and scale 
with a decreasing convective core.
 
 The third frequency of RV Ari can only be associated with non-radial mode because of its proximity to the second frequency,
 which is the first overtone radial mode. It is very likely a dipole or quadrupole mode because  of the disk averaging effect
 in photometric amplitudes.
 We made a working hypothesis that $\nu_3$ is a dipole axisymmetric mode and repeated seismic modelling. Thus, we fitted 
 the three frequencies  and the parameter $f$ for the dominant frequency.  Including  the nonradial mode constrained 
 enormously, in particular, the global  stellar parameters.   
As before, the number of seismic models in the HSB phase is much larger.
Moreover, the HSB and OC seismic models have different ranges of the rotational velocity. 
Thus, perhaps  independent information on the rotation would finally decide between these two phases of evolution.

\section*{Acknowledgements}
The work was financially supported by the Polish NCN grant 2018/29/B/ST9/02803.
Calculations have been partly carried out using resources provided by Wroclaw Centre for Networking and Supercomputing (http://www.wcss.pl), grant No. 265. This paper includes data collected by the TESS mission. Funding for the TESS mission 
is provided by the NASA's Science Mission Directorate.
This work has made use of data from the European Space Agency (ESA) mission
{\it Gaia} (\url{https://www.cosmos.esa.int/gaia}), processed by the {\it Gaia}
Data Processing and Analysis Consortium (DPAC,
\url{https://www.cosmos.esa.int/web/gaia/dpac/consortium}). Funding for the DPAC
has been provided by national institutions, in particular the institutions
participating in the {\it Gaia} Multilateral Agreement.

\section*{Data Availability}
The TESS data are available from the NASA MAST portal  https://archive.stsci.edu/.
The ASAS observations are available at the website of \url{http://www.astrouw.edu.pl/asas}.
Theoretical computations will be shared on reasonable request to the corresponding author.



\bibliographystyle{mnras}
\bibliography{JDD_biblio1} 



\appendix

\section{The frequencies from the Fourier analysis of the TESS light curves}

List of extracted frequencies for AE UMa and RV Ari  in the order they were found in the TESS data.
In Table\,A1, we give the frequencies of AE UMa obtained from the data separately  from the sector S21 and S48.
Table\,A2 contains the frequencies of RV Ari obtained from the two combined sectors, S42 and S43.
\begin{table*}
	\centering
	\caption{The whole set of frequencies found in the TESS observations of AE UMa from sector 21 and from sector 48.
		The following columns give the frequency ID, frequency value, amplitude and signal-to-noise ratio.
     	Formal errors are given in parentheses.}
	\label{tab:full_TESS_AE_UMa_s21}
	\begin{tabular}{ r r r  r r r r r } 
		\hline
		\multicolumn{7}{l}{\bf S21} - 57 significant peaks & \\
		\hline
		ID    &  $\nu\,(\mathrm{d}^{-1})$               &        A (ppt)         &   $S/N$        & ID    &  $\nu\,(\mathrm{d}^{-1})$               &        A (ppt)     &     $S/N$       \\
		\hline
		 $\nu_{1}$             &        11.625687(3)  &          131.88(2)  &           9.5      &  $6\nu_{1}+2\nu_{2}$    &          99.8153(6)  &            0.69(2)  &           8.8  \\
		$2\nu_{1}$            &        23.251413(9)  &           49.07(2)  &           9.4      &  $3\nu_{1}-2\nu_{2}$    &           4.8158(7)  &            0.64(2)  &           8.3  \\
		$\nu_{2}$             &         15.03139(1)  &           30.64(2)  &           9.4      &  $5\nu_{1}-\nu_{2}$    &          43.0954(7)  &            0.61(2)  &           8.6  \\
		$\nu_{1}+\nu_{2}$     &         26.65705(2)  &           21.36(2)  &           9.4      &  $7\nu_{1}+2\nu_{2}$    &         111.4422(7)  &            0.59(2)  &           8.6  \\
		$-\nu_{1}+\nu_{2}$    &          3.40569(2)  &           20.06(2)  &           9.4      &  $6\nu_{1}-\nu_{2}$    &          54.7204(8)  &            0.56(2)  &           7.9  \\
		$3\nu_{1}$            &         34.87682(2)  &           19.42(2)  &           9.4      &  $\nu_{1}+3\nu_{2}$    &          56.7190(8)  &            0.52(2)  &           8.2  \\
		$2\nu_{1}+\nu_{2}$    &         38.28254(4)  &           12.16(2)  &           9.4      &  $4\nu_{1}-2\nu_{2}$    &          16.4426(9)  &            0.50(2)  &           7.9  \\
		$4\nu_{1}$            &         46.50241(5)  &            9.04(2)  &           9.4      &  $9\nu_{1}+\nu_{2}$    &         119.6604(9)  &            0.49(2)  &           7.5  \\
		$3\nu_{1}+\nu_{2}$    &         49.90797(6)  &            6.87(2)  &           9.4      &  $7\nu_{1}-\nu_{2}$    &          66.3453(9)  &            0.48(2)  &           7.9  \\
		$2\nu_{1}-\nu_{2}$    &          8.22030(7)  &            6.64(2)  &           9.4      &  $3\nu_{2}$               &          45.0946(9)  &            0.46(2)  &           7.8  \\
		$5\nu_{1}$            &          58.1278(1)  &            4.26(2)  &           9.4      &  $9\nu_{1}$               &          104.629(1)  &            0.41(2)  &           7.7  \\
		$4\nu_{1}+\nu_{2}$    &          61.5335(1)  &            3.86(2)  &           9.4      &  $8\nu_{1}+2\nu_{2}$    &          123.068(1)  &            0.39(2)  &           7.5  \\
		$2\nu_{2}$            &          30.0628(1)  &            3.34(2)  &           9.4      &  $8\nu_{1}-\nu_{2}$    &           77.973(1)  &            0.38(2)  &           7.3  \\
		$\nu_{1}+2\nu_{2}$    &          41.6885(1)  &            3.30(2)  &           9.3      &  $5\nu_{1}-2\nu_{2}$    &           28.064(1)  &            0.36(2)  &           7.3  \\
		$3\nu_{1}-\nu_{2}$    &          19.8458(2)  &            2.68(2)  &           9.3      &  $10\nu_{1}+\nu_{2}$    &          131.286(1)  &            0.33(2)  &           7.2  \\
		$5\nu_{1}+\nu_{2}$    &          73.1590(2)  &            2.59(2)  &           9.2      &  $9\nu_{1}+2\nu_{2}$    &          134.694(1)  &            0.32(2)  &           6.9  \\
		$6\nu_{1}$            &          69.7533(2)  &            2.30(2)  &           9.4      &  $2\nu_{1}+3\nu_{2}$    &           68.345(1)  &            0.32(2)  &           7.4  \\
		$2\nu_{1}+2\nu_{2}$   &          53.3137(2)  &            1.99(2)  &           9.2      &  $3\nu_{1}+3\nu_{2}$    &           79.969(2)  &            0.28(2)  &           5.9  \\
		$-\nu_{1}+2\nu_{2}$   &          18.4369(2)  &            1.88(2)  &           9.4      &  $4\nu_{1}+3\nu_{2}$    &           91.595(2)  &            0.26(2)  &           6.1  \\
		$6\nu_{1}+\nu_{2}$    &          84.7848(2)  &            1.75(2)  &           9.2      &  $10\nu_{1}$               &          116.254(2)  &            0.25(2)  &           6.0  \\
		$3\nu_{1}+2\nu_{2}$   &          64.9390(3)  &            1.40(2)  &           9.0      &  $5\nu_{1}+3\nu_{2}$    &          103.221(2)  &            0.25(2)  &           6.1  \\
		$7\nu_{1}$            &          81.3789(3)  &            1.35(2)  &           9.2      &  $9\nu_{1}-\nu_{2}$    &           89.598(2)  &            0.25(2)  &           5.8  \\
		$4\nu_{1}-\nu_{2}$    &          31.4720(4)  &            1.22(2)  &           9.1      &  $-\nu_{1}+3\nu_{2}$    &           33.470(2)  &            0.22(2)  &           5.9  \\
		$7\nu_{1}+\nu_{2}$    &          96.4101(4)  &            1.15(2)  &           9.0      &  $7\nu_{1}+3\nu_{2}$    &          126.474(2)  &            0.21(2)  &           6.1  \\
		$4\nu_{1}+2\nu_{2}$   &          76.5645(4)  &            1.09(2)  &           8.9      &  $10\nu_{1}+2\nu_{2}$    &          146.317(2)  &            0.20(2)  &           6.0  \\
		$5\nu_{1}+2\nu_{2}$   &          88.1901(5)  &            0.92(2)  &           8.9      &  $11\nu_{1}+\nu_{2}$    &          142.911(2)  &            0.21(2)  &           5.5  \\
		$8\nu_{1}$            &          93.0043(5)  &            0.79(2)  &           8.5      &  $6\nu_{1}-2\nu_{2}$    &           39.694(2)  &            0.21(2)  &           5.1  \\
		$-2\nu_{1}+2\nu_{2}$  &           6.8111(6)  &            0.79(2)  &           8.8      &  $6\nu_{1}+3\nu_{2}$    &          114.844(2)  &            0.20(2)  &           5.7  \\
		$8\nu_{1}+\nu_{2}$    &         108.0344(6)  &            0.74(2)  &           8.7      &     \\
		\hline
\multicolumn{7}{l}{\bf S48} - 55 significant peaks & \\
		\hline
		ID    &  $\nu\,(\mathrm{d}^{-1})$               &        A (ppt)         &   $S/N$        & ID    &  $\nu\,(\mathrm{d}^{-1})$               &        A (ppt)     &     $S/N$       \\
		\hline
		$\nu_{1}$             &        11.625523(4)  &          131.49(3)  &           8.4  &  $6\nu_{1}+2\nu_{2}$    &          99.8167(6)  &            0.78(3)  &           8.0   \\
		$2\nu_{1}$            &         23.25104(1)  &           48.68(3)  &           8.4  &  $8\nu_{1}$               &          93.0057(6)  &            0.76(3)  &           8.1   \\
		$\nu_{2}$             &         15.03113(2)  &           30.77(3)  &           8.4  &  $3\nu_{1}-2\nu_{2}$    &           4.8139(7)  &            0.67(3)  &           7.5   \\
		$\nu_{1}+\nu_{2}$     &         26.65662(2)  &           21.42(3)  &           8.4  &  $5\nu_{1}-\nu_{2}$    &          43.0980(8)  &            0.63(3)  &           7.9   \\
		$-\nu_{1}+\nu_{2}$    &          3.40558(2)  &           19.90(3)  &           8.4  &  $7\nu_{1}+2\nu_{2}$    &         111.4416(9)  &            0.57(3)  &           7.4   \\
		$3\nu_{1}$            &         34.87683(3)  &           19.39(3)  &           8.4  &  $1\nu_{1}+3\nu_{2}$    &          56.7190(9)  &            0.55(3)  &           7.5   \\
		$2\nu_{1}+\nu_{2}$    &         38.28240(4)  &           12.26(3)  &           8.4  &  $9\nu_{1}+\nu_{2}$    &         119.6611(9)  &            0.51(3)  &           7.9   \\
		$4\nu_{1}$            &         46.50250(5)  &            9.07(3)  &           8.3  &  $6\nu_{1}-\nu_{2}$    &          54.7249(9)  &            0.51(3)  &           7.8   \\
		$3\nu_{1}+\nu_{2}$    &         49.90817(7)  &            6.93(3)  &           8.4  &  $7\nu_{1}-\nu_{2}$    &          66.3499(10)  &            0.49(3)  &           7.2   \\
		$2\nu_{1}-\nu_{2}$    &          8.21952(7)  &            6.57(3)  &           8.3  &  $3\nu_{2}$               &           45.094(1)  &            0.47(3)  &           7.1   \\
		$5\nu_{1}$            &          58.1284(1)  &            4.29(3)  &           8.4  &  $4\nu_{1}-2\nu_{2}$    &           16.438(1)  &            0.44(3)  &           7.1   \\
		$4\nu_{1}+\nu_{2}$    &          61.5339(1)  &            3.89(3)  &           8.3  &  $9\nu_{1}$               &          104.631(1)  &            0.43(3)  &           7.2   \\
		$2\nu_{2}$            &          30.0622(1)  &            3.42(3)  &           8.4  &  $8\nu_{1}+2\nu_{2}$    &          123.069(1)  &            0.41(3)  &           7.4   \\
		$\nu_{1}+2\nu_{2}$    &          41.6877(1)  &            3.36(3)  &           8.4  &  $2\nu_{1}+3\nu_{2}$    &           68.345(1)  &            0.41(3)  &           6.6   \\
		$3\nu_{1}-\nu_{2}$    &          19.8452(2)  &            2.70(3)  &           8.3  &  $8\nu_{1}-\nu_{2}$    &           77.974(1)  &            0.36(3)  &           6.9   \\
		$5\nu_{1}+\nu_{2}$    &          73.1594(2)  &            2.59(3)  &           8.4  &  $9\nu_{1}+2\nu_{2}$    &          134.691(1)  &            0.34(3)  &           6.5   \\
		$6\nu_{1}$            &          69.7543(2)  &            2.32(3)  &           8.2  &  $5\nu_{1}-2\nu_{2}$    &           28.064(1)  &            0.34(3)  &           6.5   \\
		$2\nu_{1}+2\nu_{2}$   &          53.3138(2)  &            2.03(3)  &           8.3  &  $10\nu_{1}+\nu_{2}$    &          131.290(2)  &            0.31(3)  &           6.2   \\
		$-\nu_{1}+2\nu_{2}$   &          18.4370(3)  &            1.91(3)  &           8.3  &  $-\nu_{1}+3\nu_{2}$    &           33.466(2)  &            0.29(3)  &           5.9   \\
		$6\nu_{1}+\nu_{2}$    &          84.7857(3)  &            1.75(3)  &           8.2  &  $10\nu_{1}$               &          116.258(2)  &            0.26(3)  &           6.1   \\
		$3\nu_{1}+2\nu_{2}$   &          64.9394(3)  &            1.45(3)  &           8.1  &  $3\nu_{1}+3\nu_{2}$    &           79.969(2)  &            0.26(3)  &           5.8   \\
		$7\nu_{1}$            &          81.3800(4)  &            1.36(3)  &           8.2  &  $9\nu_{1}-\nu_{2}$    &           89.599(2)  &            0.24(3)  &           5.2   \\
		$4\nu_{1}-\nu_{2}$    &          31.4702(4)  &            1.24(3)  &           8.0  &  $5\nu_{1}+3\nu_{2}$    &          103.221(2)  &            0.24(3)  &           5.5   \\
		$7\nu_{1}+\nu_{2}$    &          96.4111(4)  &            1.17(3)  &           8.2  &  $4\nu_{1}+3\nu_{2}$    &           91.595(2)  &            0.24(3)  &           5.8   \\
		$4\nu_{1}+2\nu_{2}$   &          76.5648(4)  &            1.11(3)  &           8.1  &  $10\nu_{1}+2\nu_{2}$    &          146.318(2)  &            0.22(3)  &           5.4   \\
		$5\nu_{1}+2\nu_{2}$   &          88.1910(5)  &            0.89(3)  &           7.7  &  $6\nu_{1}+3\nu_{2}$    &          114.849(2)  &            0.22(3)  &           5.2   \\
		$8\nu_{1}+\nu_{2}$    &         108.0364(6)  &            0.79(3)  &           7.6  &  $7\nu_{1}+3\nu_{2}$    &          126.469(3)  &            0.19(3)  &           5.2   \\
		$-2\nu_{1}+2\nu_{2}$  &           6.8116(6)  &            0.77(3)  &           8.1  &     \\
		\hline
	\end{tabular}
\end{table*}

\begin{table*}
	\centering
	\caption{The whole set of 135 frequencies found in the TESS observations of RV Ari from combined sectors 42 and 43.
		The following columns give the frequency ID, frequency value, amplitude and signal-to-noise ratio.
     	Formal errors are given in parentheses.}
	\label{tab:full_TESS_RV_Ari}
	\begin{tabular}{r r r r  r r r r } 
		\hline
		ID    &  $\nu\,(\mathrm{d}^{-1})$               &        A (ppt)        &   $S/N$        & ID    &  $\nu\,(\mathrm{d}^{-1})$               &        A (ppt)         &   $S/N$        \\
		\hline
		$\nu_{1}$                  &       10.737880(3)  &          128.84(2)  &        11.8  &    $3\nu_{1}+\nu_{2}-\nu_{3}$  &         32.5010(5)  &            0.68(2)  &         9.6        \\
		$2\nu_{1}$                 &       21.475767(9)  &           42.22(2)  &        11.8  &    $8\nu_{1}+2\nu_{2}$         &        113.7017(5)  &            0.68(2)  &         8.9        \\
		$\nu_{2}$                  &       13.899137(9)  &           39.63(2)  &        10.5  &    $7\nu_{1}-\nu_{2}$          &         61.2661(6)  &            0.62(2)  &         9.7        \\
		$\nu_{1}+\nu_{2}$          &        24.63702(1)  &           25.92(2)  &        11.3  &    $\nu_{1}+2\nu_{2}+\nu_{3}$  &         52.1476(6)  &            0.63(2)  &         9.0        \\
		$-\nu_{1}+\nu_{2}$         &         3.16127(1)  &           25.82(2)  &        11.1  &    $2\nu_{2}+\nu_{3}$          &         41.4103(6)  &            0.62(2)  &         9.2        \\
		$3\nu_{1}$                 &        32.21365(2)  &           16.26(2)  &        11.7  &    $2\nu_{1}+2\nu_{2}+\nu_{3}$ &         62.8861(6)  &            0.59(2)  &         9.1        \\
		$2\nu_{1}+\nu_{2}$         &        35.37489(3)  &           13.77(2)  &        10.8  &    $5\nu_{1}+3\nu_{2}$         &         95.3866(6)  &            0.59(2)  &         8.4        \\
		$\nu_{3}$                  &        13.61183(3)  &           11.61(2)  &        11.9  &    $4\nu_{1}+\nu_{2}-\nu_{3}$  &         43.2389(6)  &            0.58(2)  &         9.5        \\
		$2\nu_{1}-\nu_{2}$         &         7.57660(4)  &            9.91(2)  &        11.0  &    $4\nu_{1}+3\nu_{2}$         &         84.6488(6)  &            0.57(2)  &         7.6        \\
		$4\nu_{1}$                 &        42.95153(5)  &            7.52(2)  &        11.4  &    $-\nu_{1}+3\nu_{2}$         &         30.9597(7)  &            0.56(2)  &         7.2        \\
		$3\nu_{1}+\nu_{2}$         &        46.11283(5)  &            6.97(2)  &        10.6  &    $3\nu_{1}-\nu_{2}+\nu_{3}$  &         31.9264(7)  &            0.56(2)  &         9.6        \\
		$\nu_{1}+2\nu_{2}$         &        38.53621(6)  &            5.76(2)  &        10.3  &    $3\nu_{1}+3\nu_{2}$         &         73.9107(7)  &            0.54(2)  &         7.0        \\
		$2\nu_{2}$                 &        27.79831(6)  &            5.67(2)  &         9.7  &    $6\nu_{1}+3\nu_{2}$         &        106.1244(7)  &            0.54(2)  &         8.2        \\
		$3\nu_{1}-\nu_{2}$         &        18.31451(6)  &            5.61(2)  &        10.9  &    $8\nu_{1}$                  &         85.9029(7)  &            0.53(2)  &8.4                 \\
		$-\nu_{1}+\nu_{3}$         &         2.87400(7)  &            5.23(2)  &        11.8  &    $9\nu_{1}+\nu_{2}$          &        110.5396(7)  &            0.51(2)  &         8.6        \\
		$\nu_{1}+\nu_{3}$          &        24.34969(8)  &            4.43(2)  &        11.6  &    $2\nu_{3}$                  &         27.2230(7)  &            0.53(2)  &9.5                 \\
		$4\nu_{1}+\nu_{2}$         &        56.85068(9)  &            3.97(2)  &        10.8  &    $5\nu_{1}-\nu_{3}$          &         40.0776(7)  &            0.50(2)  &         8.7        \\
		$5\nu_{1}$                 &         53.6895(1)  &            3.49(2)  &        10.9  &    $7\nu_{1}+3\nu_{2}$         &        116.8627(7)  &            0.49(2)  &         7.8        \\
		$4\nu_{1}-\nu_{2}$         &         29.0525(1)  &            3.38(2)  &        10.6  &    $\nu_{1}+2\nu_{3}$          &         37.9620(8)  &            0.48(2)  &         9.1        \\
		$2\nu_{1}+\nu_{3}$         &         35.0876(1)  &            3.37(2)  &        11.8  &    $9\nu_{1}+2\nu_{2}$         &        124.4400(8)  &            0.47(2)  &         8.3        \\
		$-\nu_{1}+2\nu_{2}$        &         17.0603(1)  &            3.34(2)  &         9.7  &    $4\nu_{1}-\nu_{2}+\nu_{3}$  &         42.6646(8)  &            0.47(2)  &         8.9        \\
		$2\nu_{1}+2\nu_{2}$        &         49.2740(1)  &            2.85(2)  &         9.1  &    $3\nu_{1}+2\nu_{2}+\nu_{3}$ &         73.6236(8)  &            0.45(2)  &         8.6        \\
		$5\nu_{1}+\nu_{2}$         &         67.5886(1)  &            2.74(2)  &        11.2  &    $-2\nu_{1}+\nu_{2}+\nu_{3}$ &          6.0349(9)  &            0.43(2)  &         8.6        \\
		$\nu_{2}+\nu_{3}$          &         27.5109(1)  &            2.52(2)  &        10.8  &    $5\nu_{1}+\nu_{2}+\nu_{3}$  &         81.1997(8)  &            0.44(2)  &         9.6        \\
		$2\nu_{1}-\nu_{3}$         &          7.8641(2)  &            2.17(2)  &        11.5  &    $5\nu_{1}-\nu_{2}+\nu_{3}$  &         53.4028(9)  &            0.43(2)  &         8.4        \\
		$5\nu_{1}-\nu_{2}$         &         39.7904(2)  &            2.02(2)  &        10.6  &    $-\nu_{1}+2\nu_{2}+\nu_{3}$ &         30.6718(9)  &            0.40(2)  &         7.8        \\
		$3\nu_{1}+2\nu_{2}$        &         60.0120(2)  &            1.97(2)  &         8.9  &    $5\nu_{1}+\nu_{2}-\nu_{3}$  &         53.9774(9)  &            0.40(2)  &         8.1        \\
		$3\nu_{1}+\nu_{3}$         &         45.8255(2)  &            1.95(2)  &        11.8  &    $5\nu_{1}-2\nu_{2}$         &         25.8919(9)  &            0.40(2)  &         7.2        \\
		$\nu_{1}+\nu_{2}+\nu_{3}$  &         38.2489(2)  &            1.87(2)  &        10.5  &    $4\nu_{1}-2\nu_{2}$         &         15.1540(9)  &            0.39(2)  &         7.2        \\
		$6\nu_{1}+\nu_{2}$         &         78.3264(2)  &            1.82(2)  &        11.1  &    $8\nu_{1}+3\nu_{2}$         &       127.6006(10)  &            0.38(2)  &         7.7        \\
		$6\nu_{1}$                 &         64.4274(2)  &            1.77(2)  &        10.2  &    $5\nu_{1}+\nu_{3}$          &        67.3016(10)  &            0.38(2)  &         8.5        \\
		$4\nu_{1}+2\nu_{2}$        &         70.7499(2)  &            1.59(2)  &         9.7  &    $4\nu_{1}+2\nu_{2}+\nu_{3}$ &          84.361(1)  &            0.36(2)  &         8.2        \\
		$2\nu_{1}+\nu_{2}+\nu_{3}$ &         48.9868(2)  &            1.60(2)  &        10.6  &    $8\nu_{1}-\nu_{2}$          &          72.004(1)  &            0.35(2)  &         7.7        \\
		$-\nu_{1}+\nu_{2}+\nu_{3}$ &         16.7731(2)  &            1.52(2)  &        11.1  &    $6\nu_{1}+\nu_{2}+\nu_{3}$  &          91.938(1)  &            0.35(2)  &         8.1        \\
		$-2\nu_{1}+2\nu_{2}$       &          6.3228(3)  &            1.32(2)  &        10.1  &    $6\nu_{1}-2\nu_{2}$         &          36.629(1)  &            0.34(2)  &         6.9        \\
		$5\nu_{1}+2\nu_{2}$        &         81.4876(3)  &            1.37(2)  &        10.0  &    $9\nu_{1}$                  &          96.642(1)  &            0.32(2)  &7.0                 \\
		$3\nu_{1}-\nu_{3}$         &         18.6017(3)  &            1.31(2)  &        11.1  &    $6\nu_{1}-\nu_{2}+\nu_{3}$  &          64.140(1)  &            0.32(2)  &         7.2        \\
		$\nu_{spurious}$           &          0.0633(3)  &            1.37(2)  &         5.0  &    $-2\nu_{1}+3\nu_{2}$        &          20.222(1)  &            0.31(2)  &         6.6        \\
		$7\nu_{1}+\nu_{2}$         &         89.0646(3)  &            1.20(2)  &        10.5  &    $6\nu_{1}-\nu_{3}$          &          50.816(1)  &            0.32(2)  &         6.8        \\
		$6\nu_{1}+2\nu_{2}$        &         92.2254(3)  &            1.19(2)  &        10.1  &    $2\nu_{1}+2\nu_{2}-\nu_{3}$ &          35.664(1)  &            0.32(2)  &         7.5        \\
		$6\nu_{1}-\nu_{2}$         &         50.5285(3)  &            1.18(2)  &         9.9  &    $2\nu_{1}+2\nu_{3}$         &          48.698(1)  &            0.31(2)  &         7.0        \\
		$\nu_{1}+3\nu_{2}$         &         52.4349(3)  &            1.22(2)  &         8.9  &    $3\nu_{1}+2\nu_{2}-\nu_{3}$ &          46.400(1)  &            0.31(2)  &         6.7        \\
		$\nu_{1}-\nu_{2}+\nu_{3}$  &         10.4502(3)  &            1.16(2)  &         8.8  &   $10\nu_{1}+\nu_{2}$          &         121.279(1)  &            0.30(2)  &         7.6        \\
		$3\nu_{1}+\nu_{2}+\nu_{3}$ &         59.7243(3)  &            1.14(2)  &        10.4  &   $3\nu_{1}-\nu_{2}-\nu_{3}$   &           4.702(1)  &            0.29(2)  &         7.4        \\
		$7\nu_{1}$                 &         75.1653(4)  &            0.92(2)  &        10.0  &   $7\nu_{1}-2\nu_{2}$          &          47.366(1)  &            0.28(2)  &         6.5        \\
		$7\nu_{1}+2\nu_{2}$        &        102.9636(4)  &            0.92(2)  &         9.8  &   $6\nu_{1}+\nu_{2}-\nu_{3}$   &          64.714(1)  &            0.28(2)  &         6.8        \\
		$2\nu_{1}-\nu_{2}+\nu_{3}$ &         21.1885(4)  &            0.88(2)  &         9.6  &   $9\nu_{1}+3\nu_{2}$          &         138.337(1)  &            0.29(2)  &         6.5        \\
		$4\nu_{1}+\nu_{3}$         &         56.5627(4)  &            0.86(2)  &        10.2  &   $10\nu_{1}+2\nu_{2}$         &         135.176(1)  &            0.28(2)  &         7.0        \\
		$4\nu_{1}-\nu_{3}$         &         29.3397(4)  &            0.84(2)  &        11.1  &   $4\nu_{1}+2\nu_{2}-\nu_{3}$  &          57.139(1)  &            0.28(2)  &         7.0        \\
		$3\nu_{2}$                 &         41.6977(4)  &            0.84(2)  &         7.7  &   $5\nu_{1}+2\nu_{2}+\nu_3$    &          95.100(1)  &            0.26(2)  &         7.1        \\
		$2\nu_{1}+3\nu_{2}$        &         63.1732(5)  &            0.81(2)  &         7.6  &   $7\nu_{1}+\nu_{2}+\nu_{3}$   &         102.676(1)  &            0.26(2)  &         6.3        \\
		$8\nu_{1}+\nu_{2}$         &         99.8021(5)  &            0.77(2)  &         9.9  &   $\nu_{1}+2\nu_{2}-\nu_{3}$   &          24.924(1)  &            0.25(2)  &         5.9        \\
		$\nu_{1}+\nu_{2}-\nu_{3}$  &         11.0251(5)  &            0.76(2)  &         7.3  &   $6\nu_{1}+4\nu_{2}$          &         120.024(1)  &            0.26(2)  &         5.6        \\
		$2\nu_{1}+\nu_{2}-\nu_{3}$ &         21.7632(5)  &            0.76(2)  &         9.9  &   $\nu_{1}+3\nu_{2}+\nu_{3}$   &          66.047(1)  &            0.25(2)  &         5.2        \\
		$3\nu_{1}-2\nu_{2}  $      &          4.4152(5)  &            0.71(2)  &         8.7  &   $\nu_{1}+4\nu_{2}$           &          66.335(1)  &            0.26(2)  &         7.0        \\
		$4\nu_{1}+\nu_{2}+\nu_{3}$ &         70.4626(5)  &            0.69(2)  &        10.2  &   $7\nu_{1}+2\nu_{2}+\nu_{3}$  &         116.576(1)  &            0.25(2)  &         6.3        \\
		\hline
	\end{tabular}
\end{table*}
\begin{table*}
	\centering
	\contcaption{}
	\label{tab:RV_Aricd}
	\begin{tabular}{r  r r r r r r r  } 
		\hline
		ID    &  $\nu\,(\mathrm{d}^{-1})$               &        A (ppt)          &   $S/N$        & ID    &  $\nu\,(\mathrm{d}^{-1})$               &        A (ppt)         &   $S/N$        \\
		\hline
		$6\nu_{1}+2\nu_{2}+\nu_{3}$&         105.839(2)  &            0.24(2)  &         6.6  &   $-\nu_{1}+2\nu_{3}$          &          16.488(2)  &            0.19(2)  &         5.5        \\
		$2\nu_{2}-\nu_{3}$         &          14.185(2)  &            0.23(2)  &         5.7  &   $7\nu_{1}+4\nu_{2}$          &         130.763(2)  &            0.19(2)  &         5.4        \\
		$4\nu_{2}$                 &          55.594(2)  &            0.22(2)  &         5.3  &   $-2\nu_{1}+2\nu_{2}+\nu_{3}$ &          19.937(2)  &            0.18(2)  &         5.7        \\
		$2\nu_{1}+4\nu_{2}$        &          77.072(2)  &            0.22(2)  &         5.2  &   $\nu_{1}+\nu_{2}+2\nu_{3}$   &          51.859(2)  &            0.19(2)  &         6.2        \\
		$5\nu_{1}+4\nu_{2}$        &         109.288(2)  &            0.22(2)  &         5.2  &   $11\nu_{1}+\nu_{2}$          &         132.016(2)  &            0.18(2)  &         5.8        \\
		$6\nu_{1}+\nu_{3}$         &          78.037(2)  &            0.21(2)  &         5.9  &   $8\nu_{1}+4\nu_{2}$          &         141.501(2)  &            0.18(2)  &         5.4        \\
		$8\nu_{1}+\nu_{2}+\nu_{3}$ &         113.416(2)  &            0.21(2)  &         6.4  &   $5\nu_{1}+2\nu_{2}-\nu_{3}$  &          67.875(2)  &            0.18(2)  &         6.0        \\
		$5\nu_{1}-\nu_{2}-\nu_{3}$ &          26.176(2)  &            0.21(2)  &         5.5  &   $2\nu_{1}+\nu_{2}+2\nu_{3}$  &          62.598(2)  &            0.17(2)  &         5.2        \\
		$4\nu_{1}+4\nu_{2}$        &          98.547(2)  &            0.20(2)  &         5.3  &   $9\nu_{1}-\nu_{2}$           &          82.743(2)  &            0.16(2)  &         5.4        \\
		$10\nu_{1}+3\nu_{2}$       &         149.077(2)  &            0.20(2)  &         5.9  &   $7\nu_{1}-\nu_{3}$           &          61.554(2)  &            0.15(2)  &         5.3        \\
		$\nu_{2}+2\nu_{3}$         &          41.123(2)  &            0.20(2)  &         5.9  &   $7\nu_{1}+\nu_{3}$           &          88.773(3)  &            0.14(2)  &         5.0        \\
		$3\nu_{1}+2\nu_{3}$        &          59.436(2)  &            0.19(2)  &         5.3  &                                &                     &                     &                    \\
		\hline
	\end{tabular}
\end{table*}

\section{Seismic models from the Bayesian analysis based on Monte Carlo simulations}

To perform seismic modelling with the most accurate sampling of various parameters, we used Bayesian analysis based 
on Monte Carlo simulations. 
This analysis was based on the Gaussian likelihood function defined as \citep[e.g.,][]{2005A&A...436..127J, 2006A&A...458..609D, 2017MNRAS.467.1433R,Jiang2021}
\begin{equation}
	{\cal L}(E|{\mathbf H})=\prod_{i=1}^n \frac1{\sqrt{2\pi\sigma_i^2}} \cdot
	{\rm exp} \left( - \frac{  ({\cal O}_i-{\cal M}_i)^2}{2\sigma_i^2}  \right),
\end{equation}
where ${\mathbf H}$ is the hypothesis that represents adjustable parameters, i.e.,
mass $M$, initial hydrogen abundance $X_0$, metallicity $Z$, initial rotational velocity $V_{\rm rot,0}$,
the mixing length parameter $\alpha_{\rm MLT}$ and convective overshooting parameter $\alpha_{\rm ov}$.
The evidence $E$ represents the calculated observables ${\cal M}_i$, i.e.,  the effective temperature $T_{\rm eff}$, luminosity $L/L_{\odot}$, pulsational frequencies, $\nu_1$ and $\nu_2$, and the non-adiabatic parameter $f$ for the dominant mode.
that can be directly compared with the observed parameters ${\cal O}_i$ determined with the errors $\sigma_i$.

All seismic  models of both stars reproduce the exact value of the dominant frequency as the radial fundamental mode.
The second frequency, corresponding  the first radial overtone,  is reproduced with an accuracy of at least 0.0005\,d$^{-1}$, 
which is approximately equivalent to numerical accuracy.  The empirical values of $f$, both its real and imaginary part,
for the dominant mode is reproduced at least within the $3\sigma$ errors.    
With these criteria, the vast majority of  seismic models, that have the values of $T_{\rm eff}, L/{\rm L}_{\sun}$
consistent with the observational values, are in the phase of hydrogen-shell burning (HSB).
In Fig.\,B1,  we show the corner plots for the parameters of the HSB seismic models of AE UMa. The corresponding histograms 
are presented in Fig.\,B2. The histograms were normalised to 1.0, so the numbers on the Y-axis times 100 are the percentage 
of models with a given parameter range.  Figs.\,B3 and B4 show the same plots for the parameters of the HSB seismic models 
of RV Ari. 

As we mentioned in the main text, for RV Ari we made the working hypothesis that its third frequency $\nu_3=13.6118$\,d$^{-1}$
is a dipole axisymmetric mode. We constructed seismic models that reproduce the $\nu_3$ with an accuracy of  0.0005\,d$^{-1}$.
In the case of HSB models our simulations converged to the two well-constrained solutions 
in ($M, X_0, Z$). This dichotomy is best seen for metallicity $Z$ what is shown in Fig.\,B5, where the values of $Z$ 
are plotted as a function of the model number. Bluish colours correspond to the models that converge to a higher $Z$ solution, 
while greenish colours to the models that converge to a lower $Z$ solution,

\begin{figure*}
	\centering
	\includegraphics[clip,width=\linewidth,height=20cm]{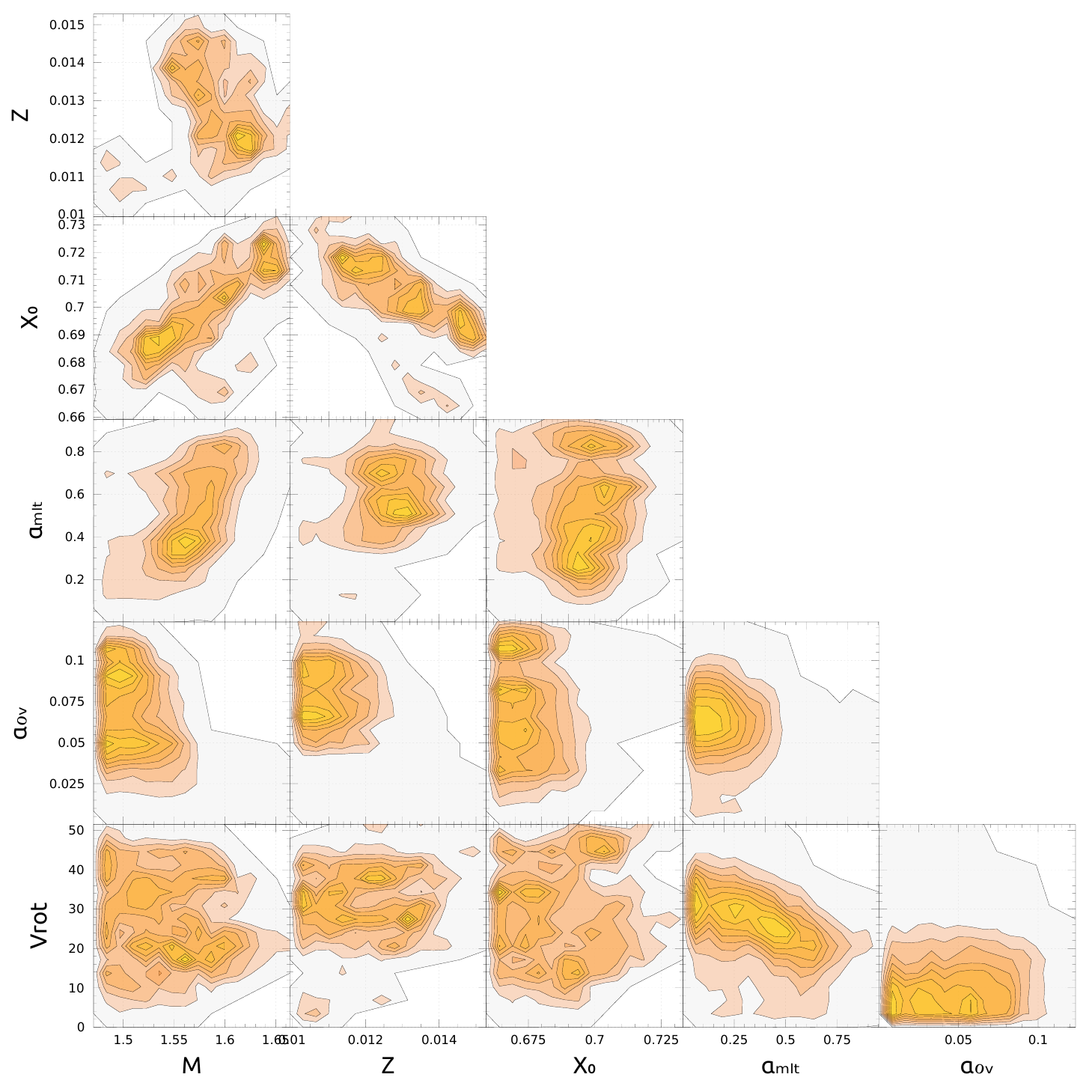}
	\caption{The corner plots for parameters obtained from complex seismic modeling of the star AE UMa.
		All models are in the hydrogen-shell burning phases of evolution and were computed	with the OPAL opacities.}
\end{figure*}

\begin{figure*}
	\centering
	\includegraphics[clip,width=0.43\linewidth,height=58mm]{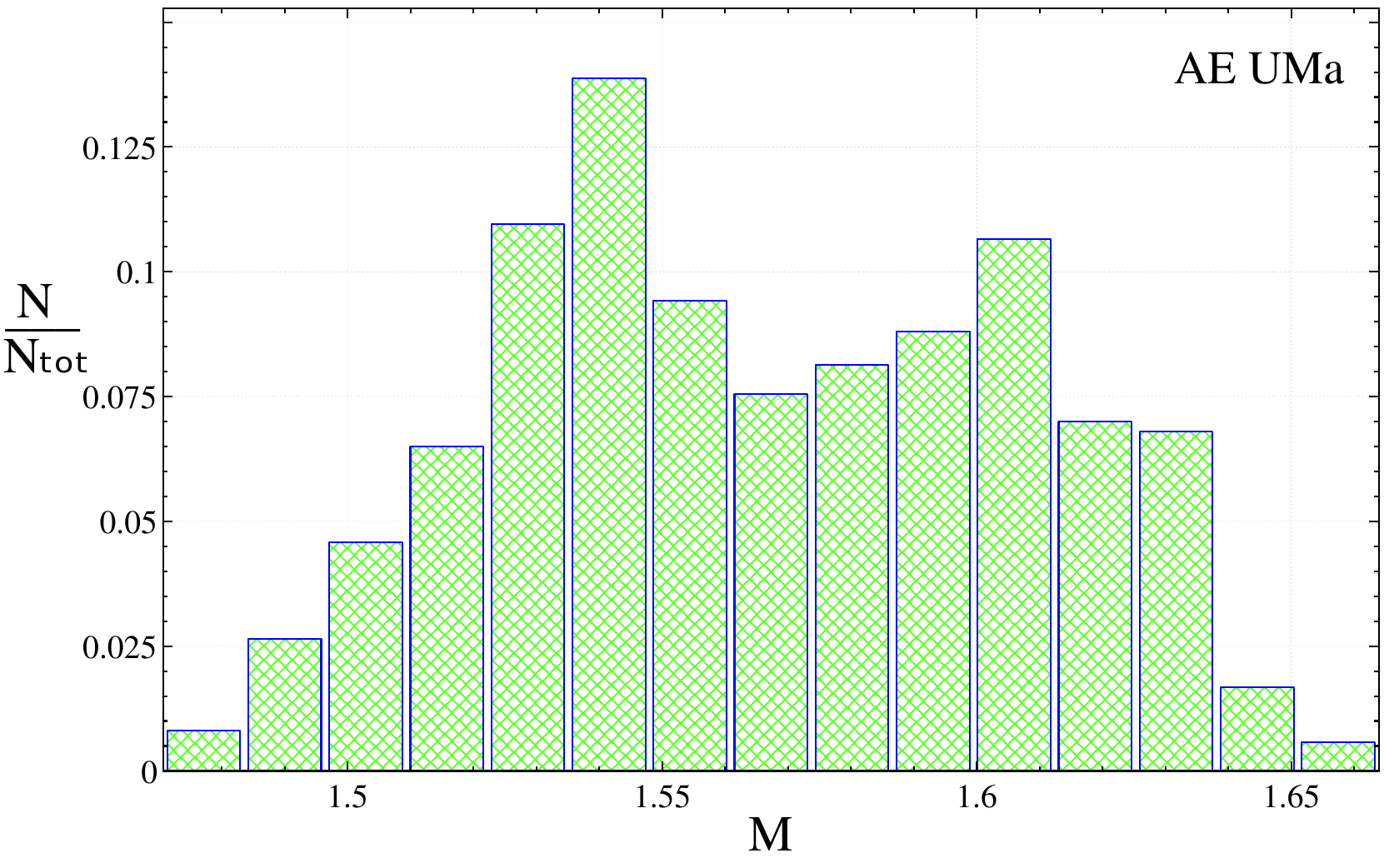}
	\includegraphics[clip,width=0.43\linewidth,height=58mm]{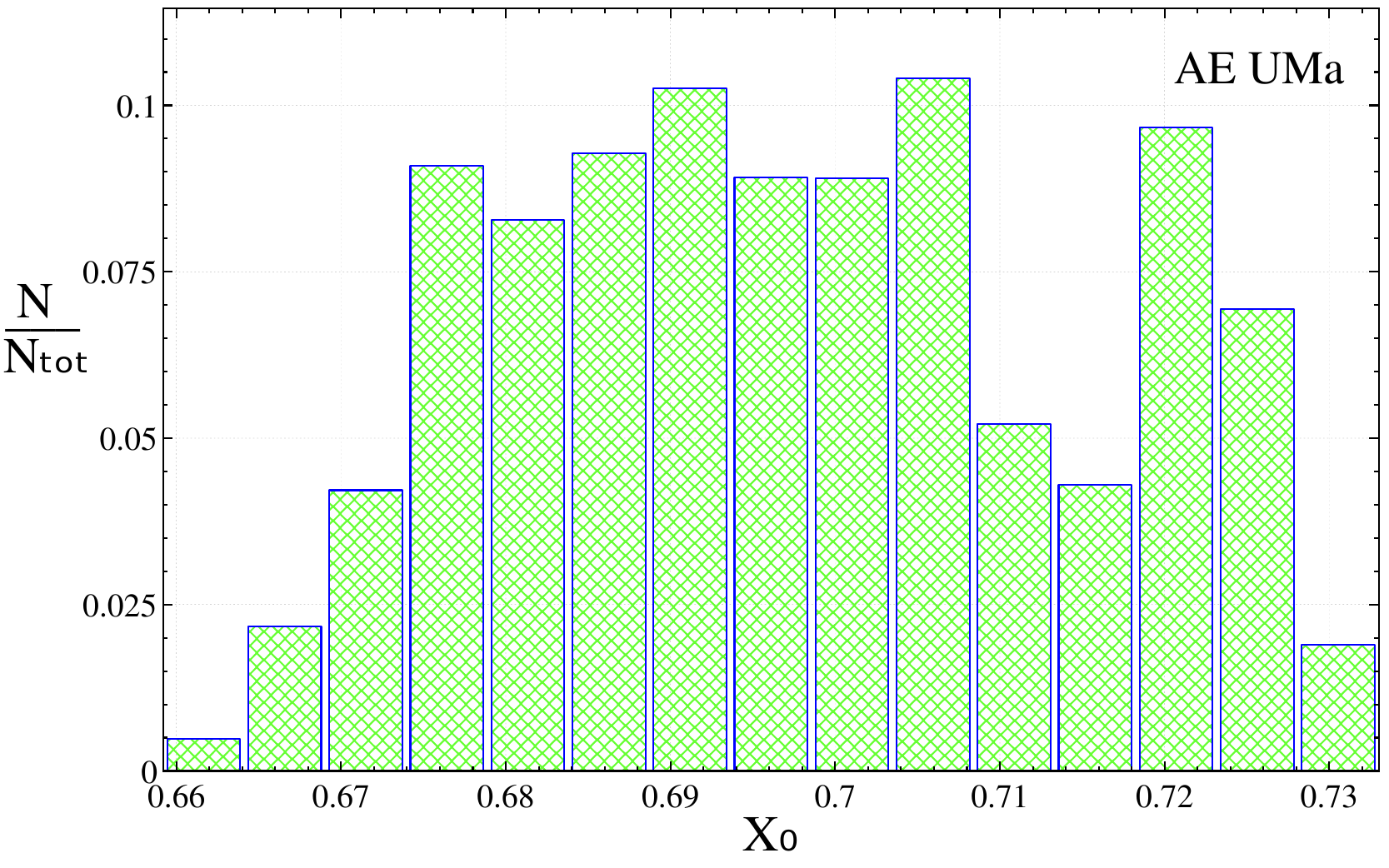}
	\includegraphics[clip,width=0.43\linewidth,height=58mm]{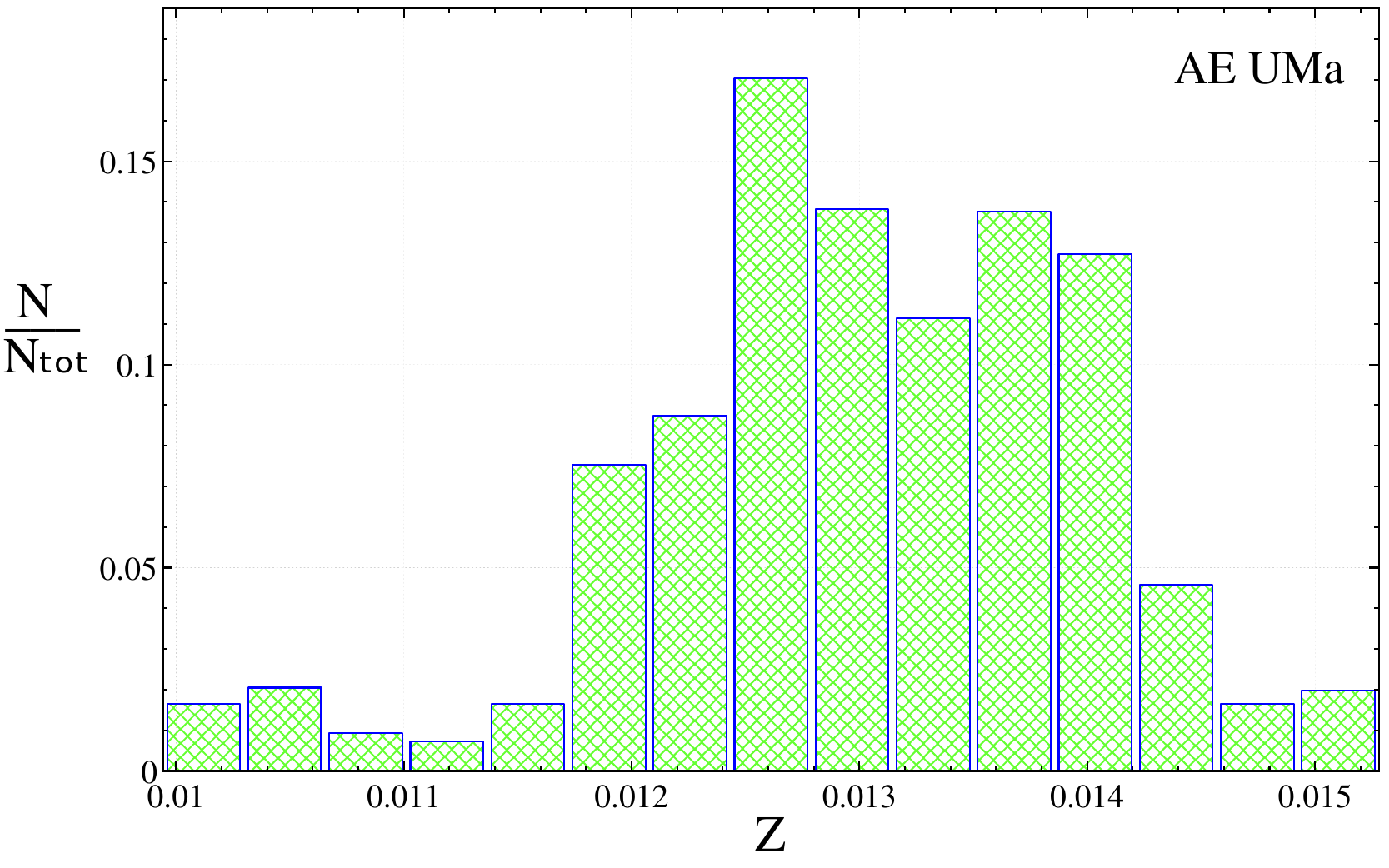}
	\includegraphics[clip,width=0.43\linewidth,height=58mm]{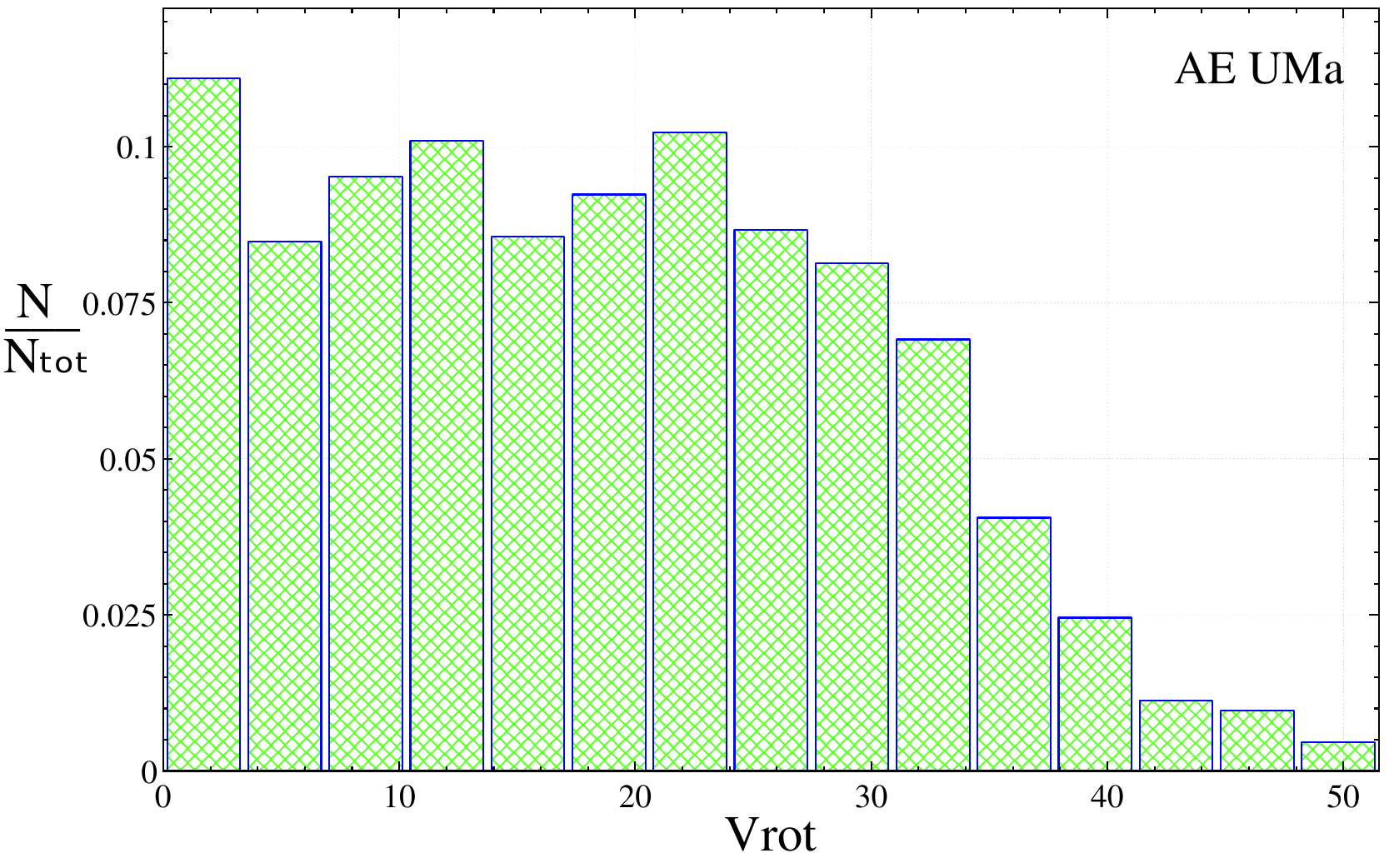}
	\includegraphics[clip,width=0.43\linewidth,height=58mm]{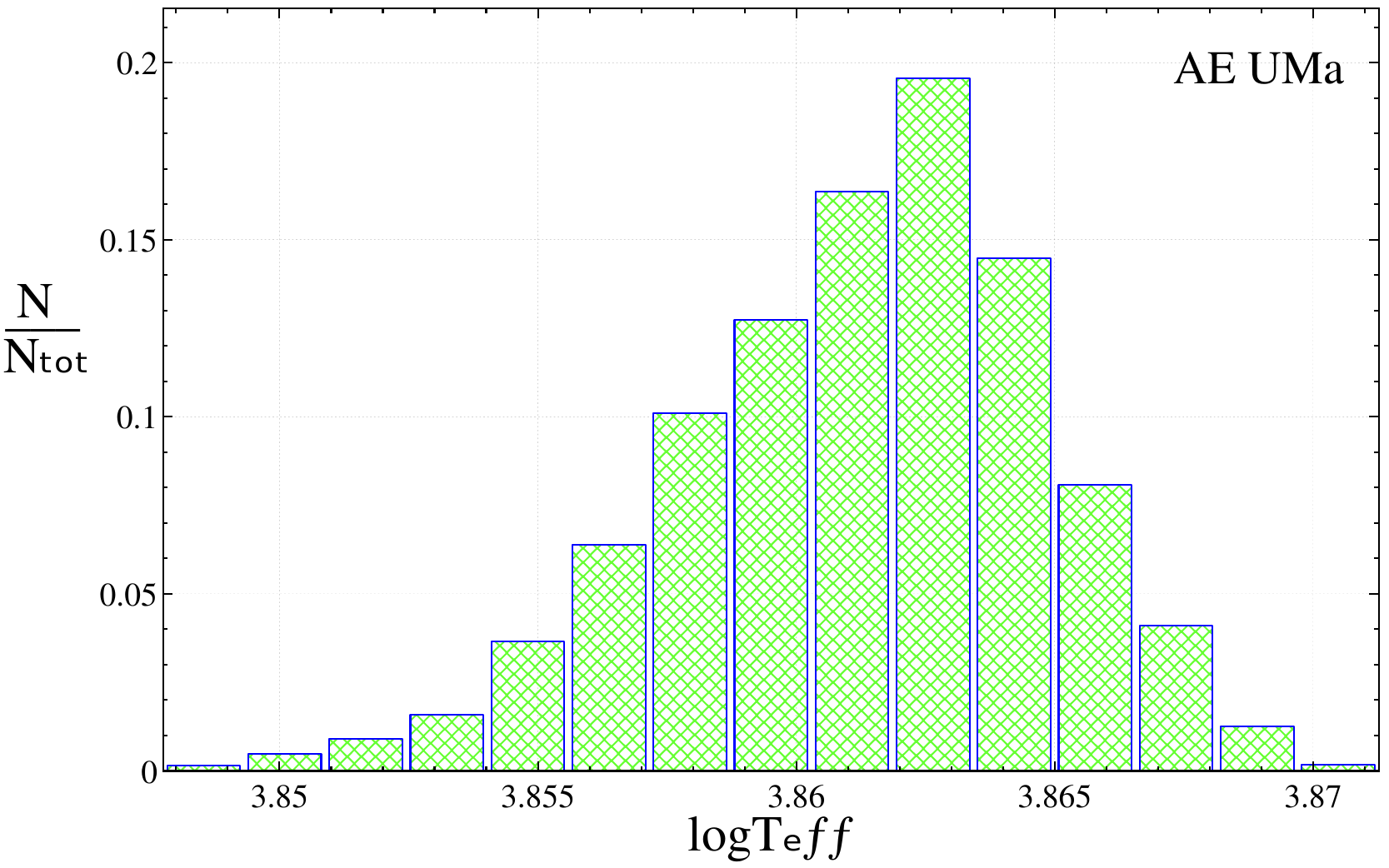}
	\includegraphics[clip,width=0.43\linewidth,height=58mm]{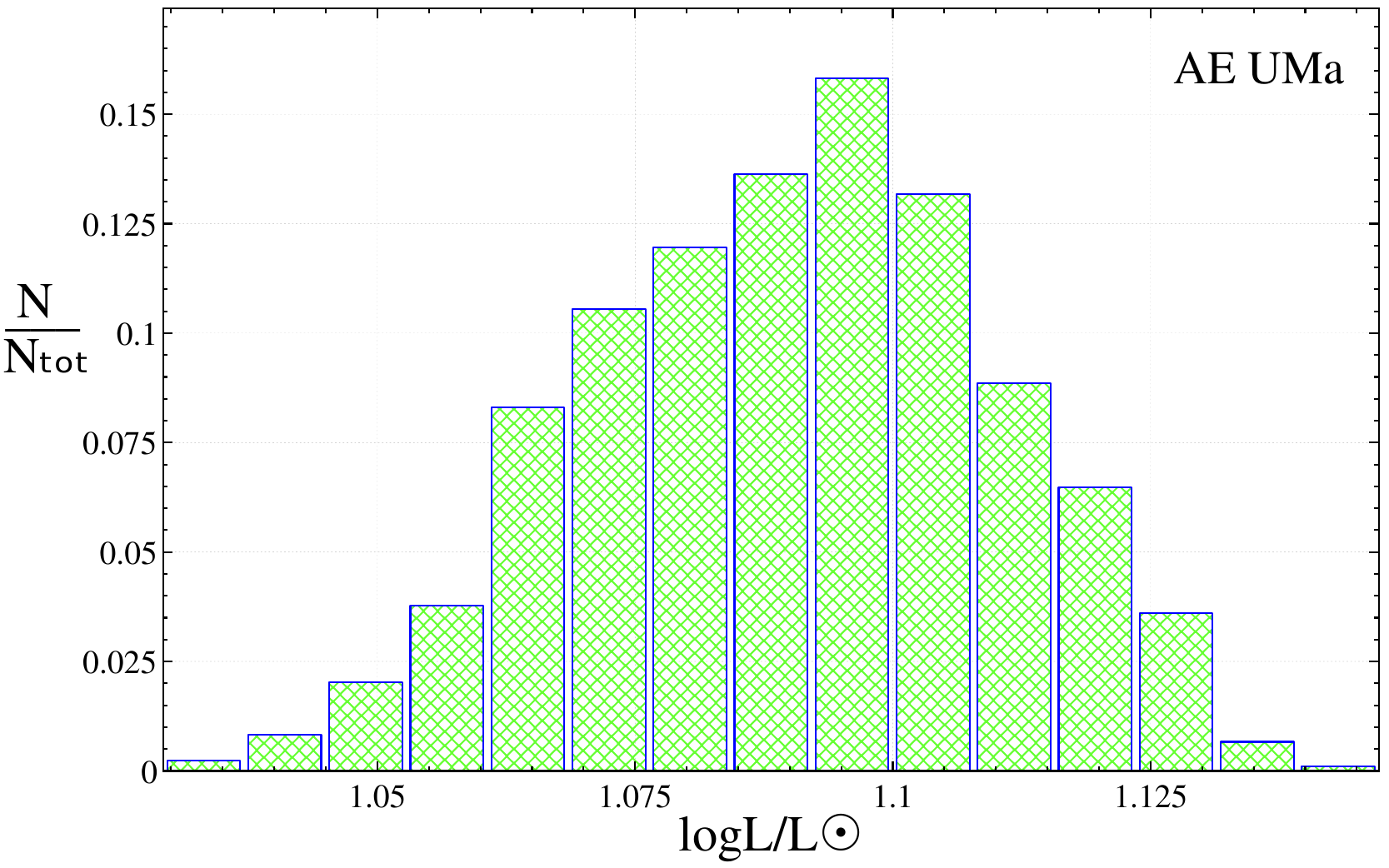}
	\includegraphics[clip,width=0.43\linewidth,height=58mm]{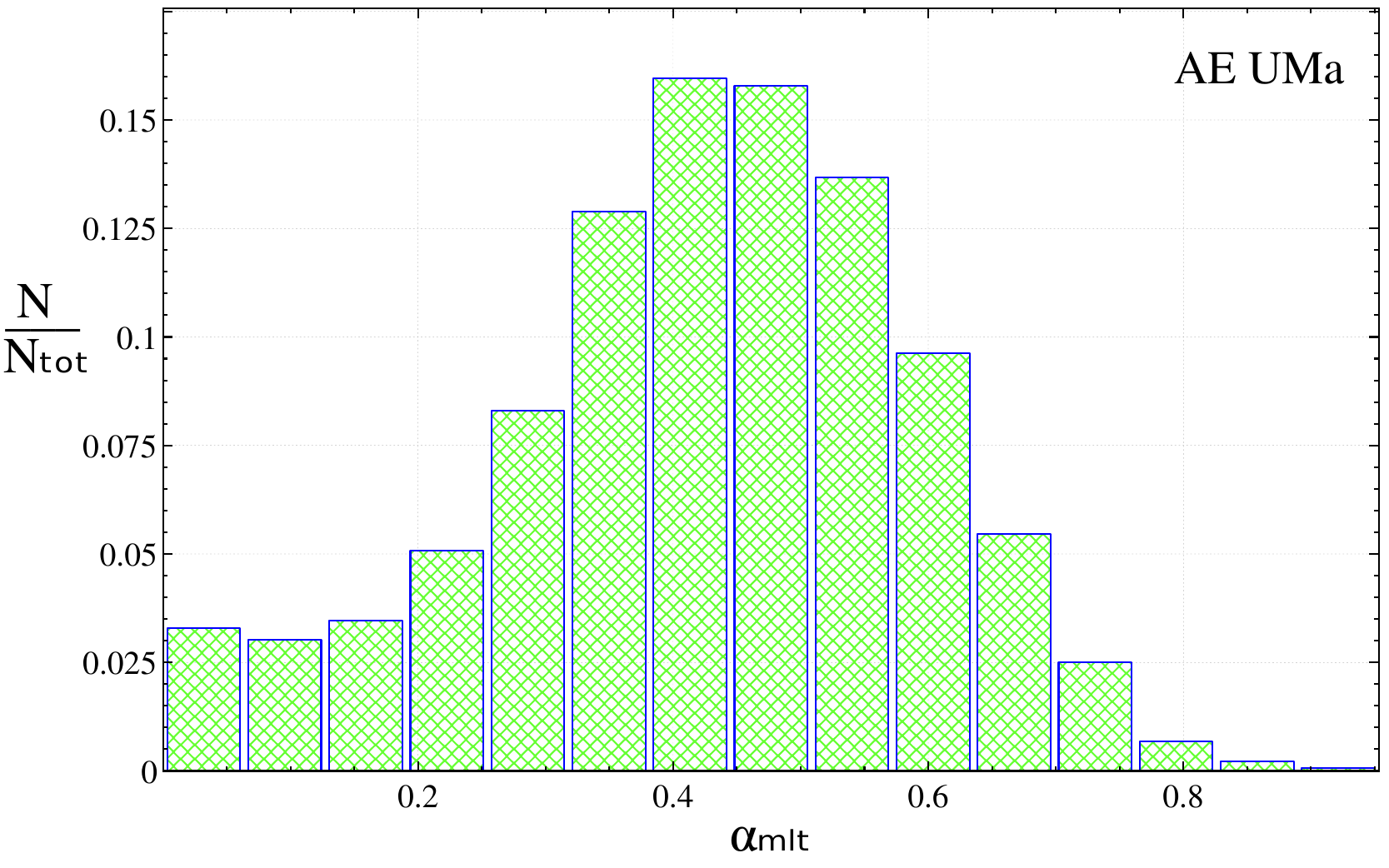}
	\includegraphics[clip,width=0.43\linewidth,height=58mm]{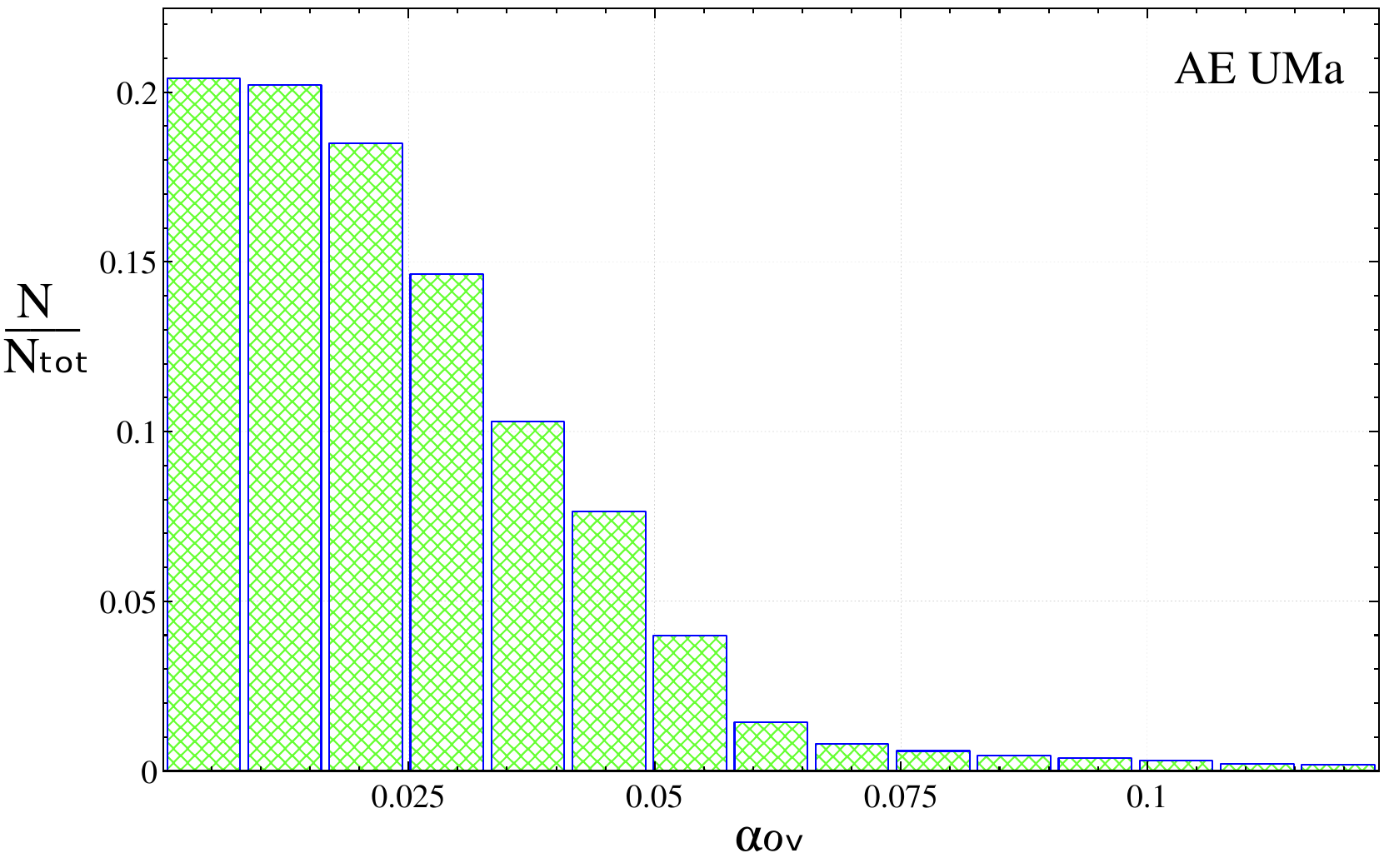}
	\caption{The corresponding histograms for determined parameters of AE UMa.}
\end{figure*}

\begin{figure*}
	\centering
	\includegraphics[clip,width=\linewidth,height=20cm]{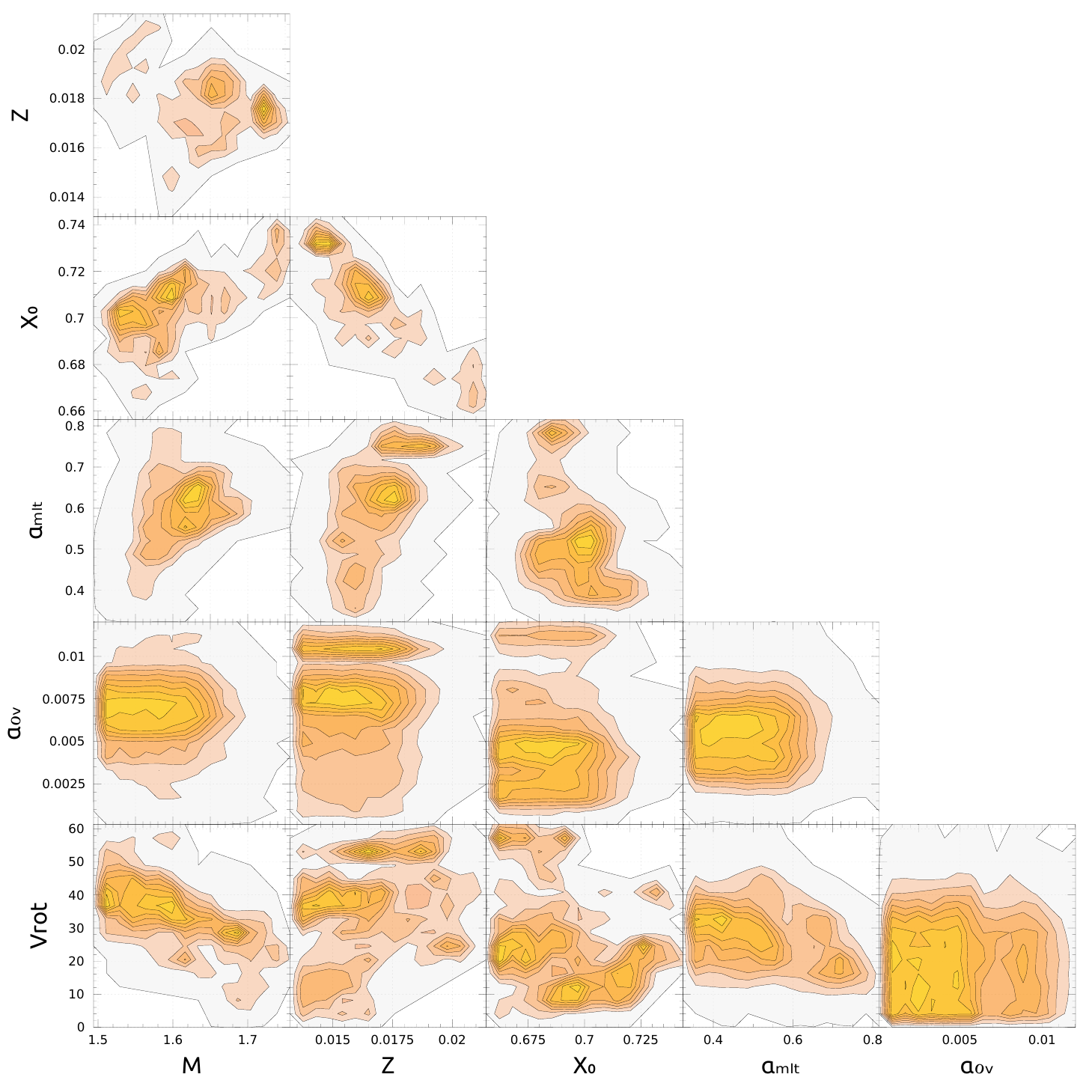}
	\caption{The corner plots for parameters obtained from complex seismic modeling of the star RV Ari.
	All models are in the hydrogen-shell burning phases of evolution and were computed	with the OPAL opacities.}
\end{figure*}

\begin{figure*}
	\centering
	\includegraphics[clip,width=0.43\linewidth,height=58mm]{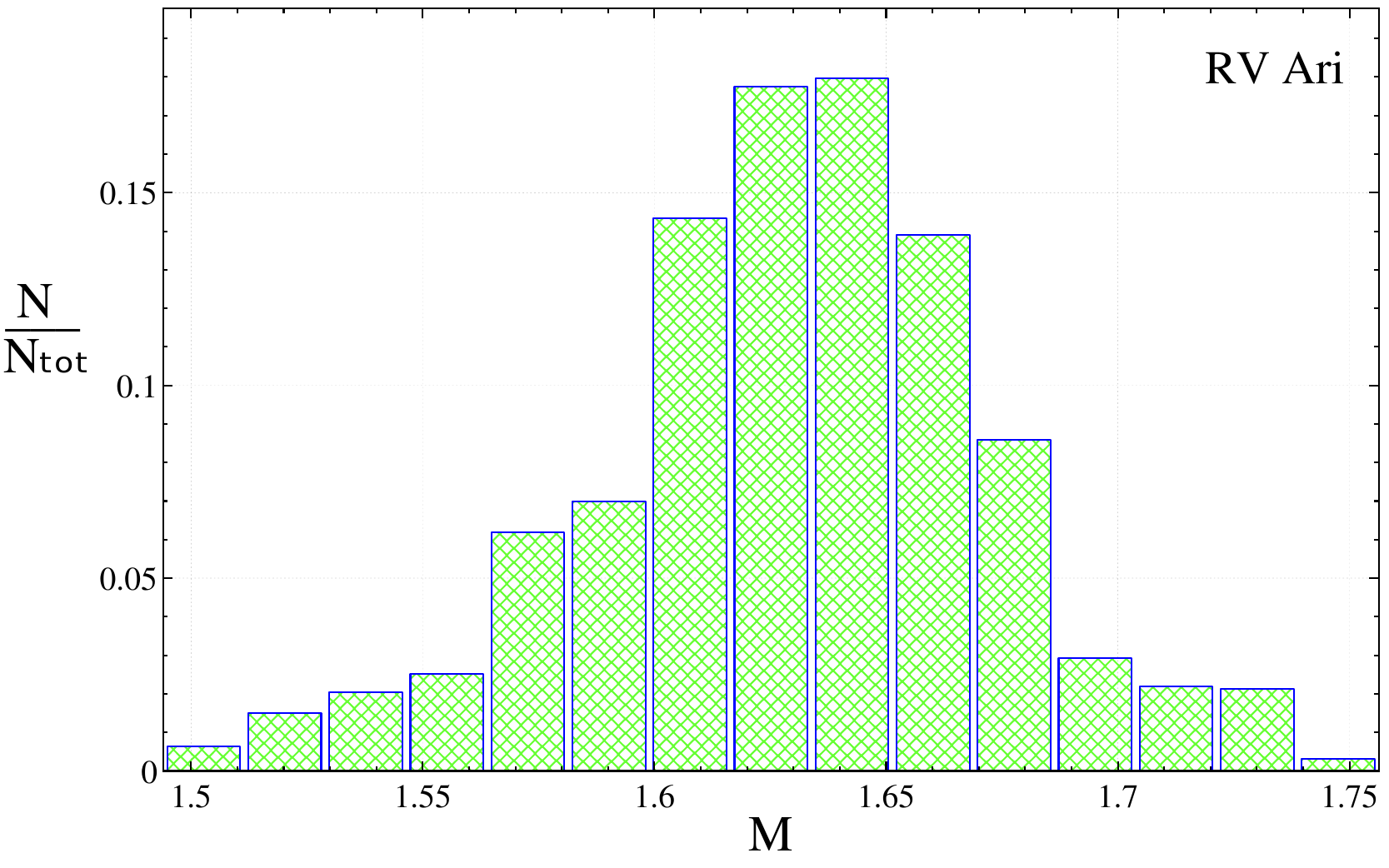}
	\includegraphics[clip,width=0.43\linewidth,height=58mm]{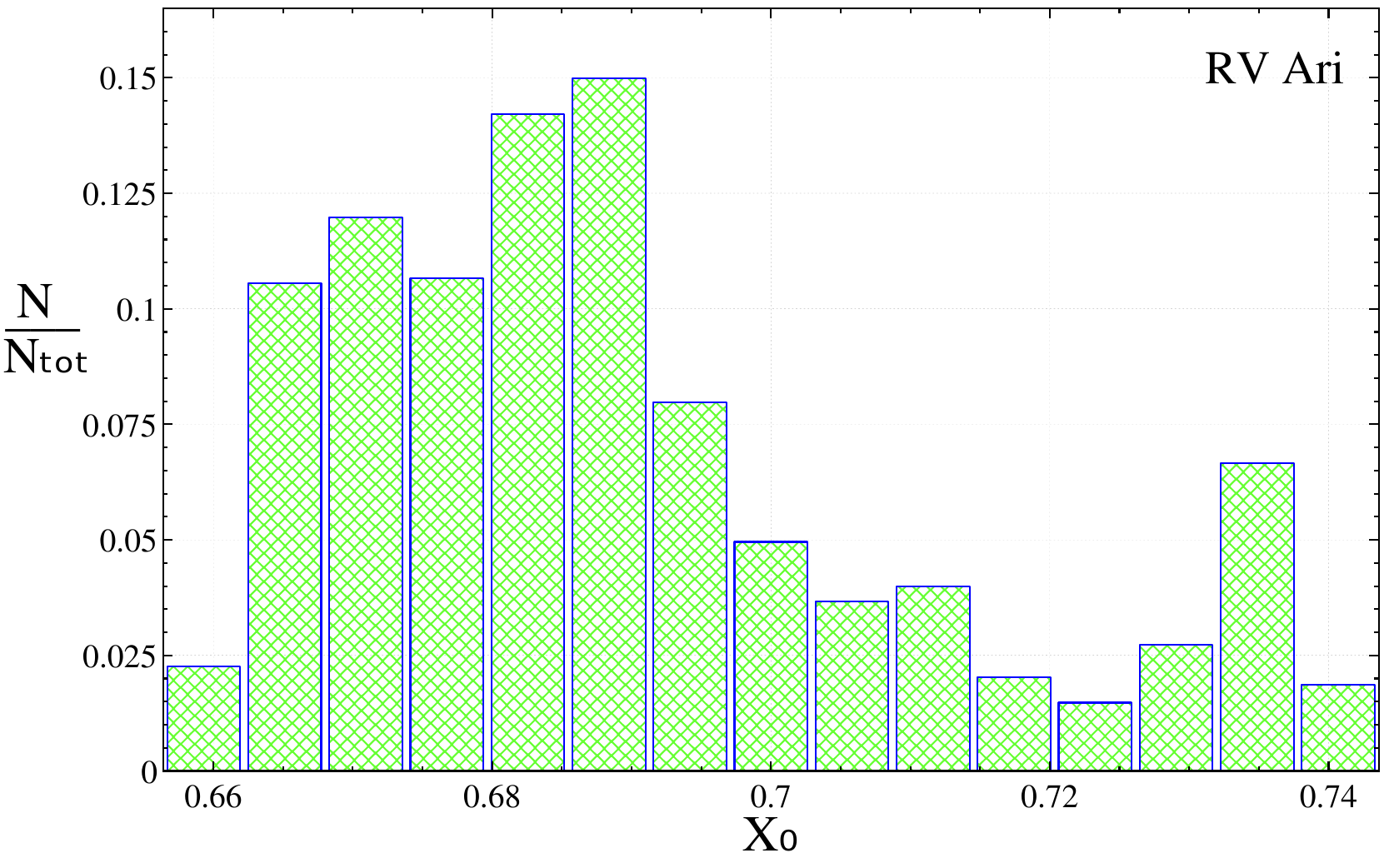}
	\includegraphics[clip,width=0.43\linewidth,height=58mm]{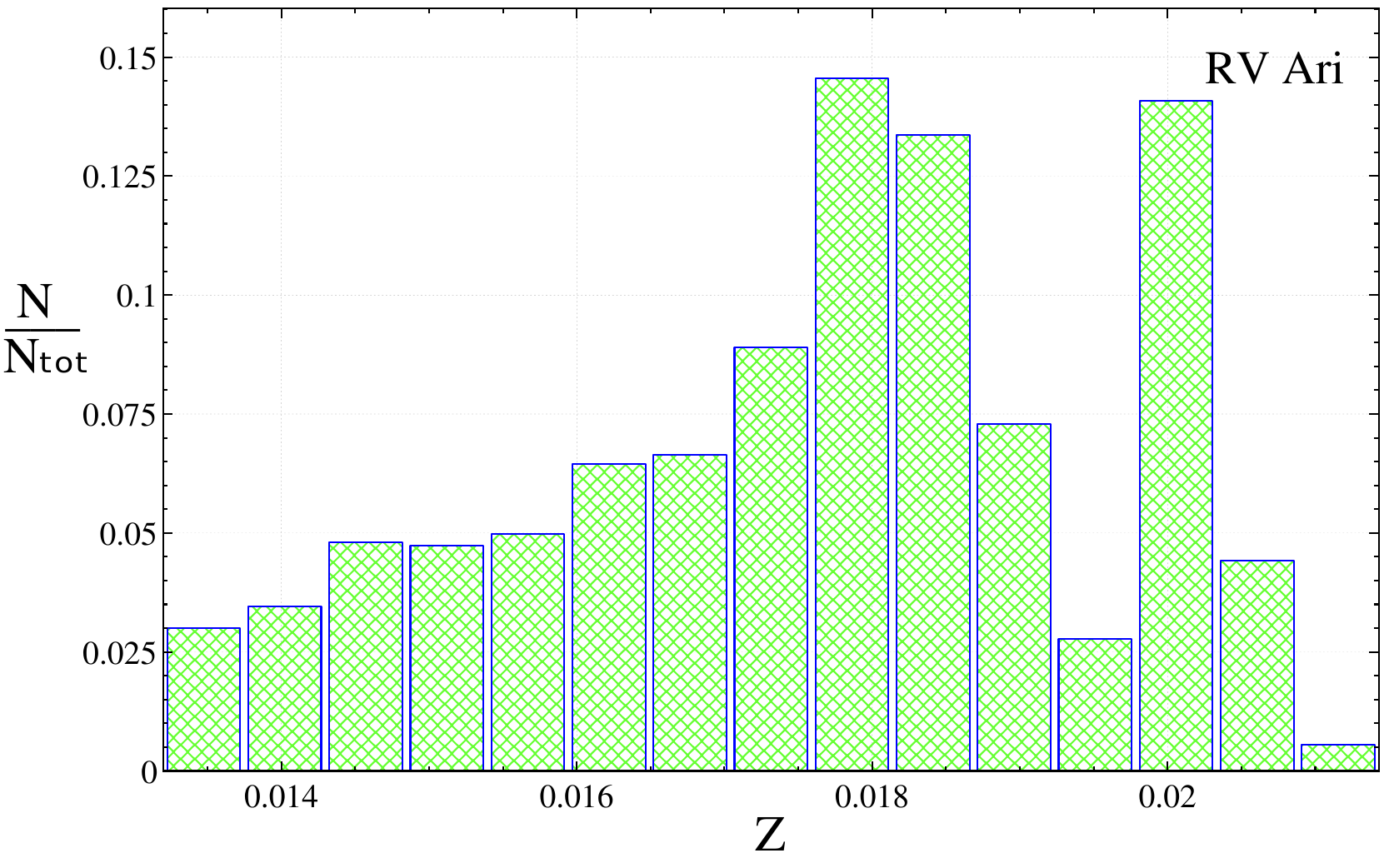}
	\includegraphics[clip,width=0.43\linewidth,height=58mm]{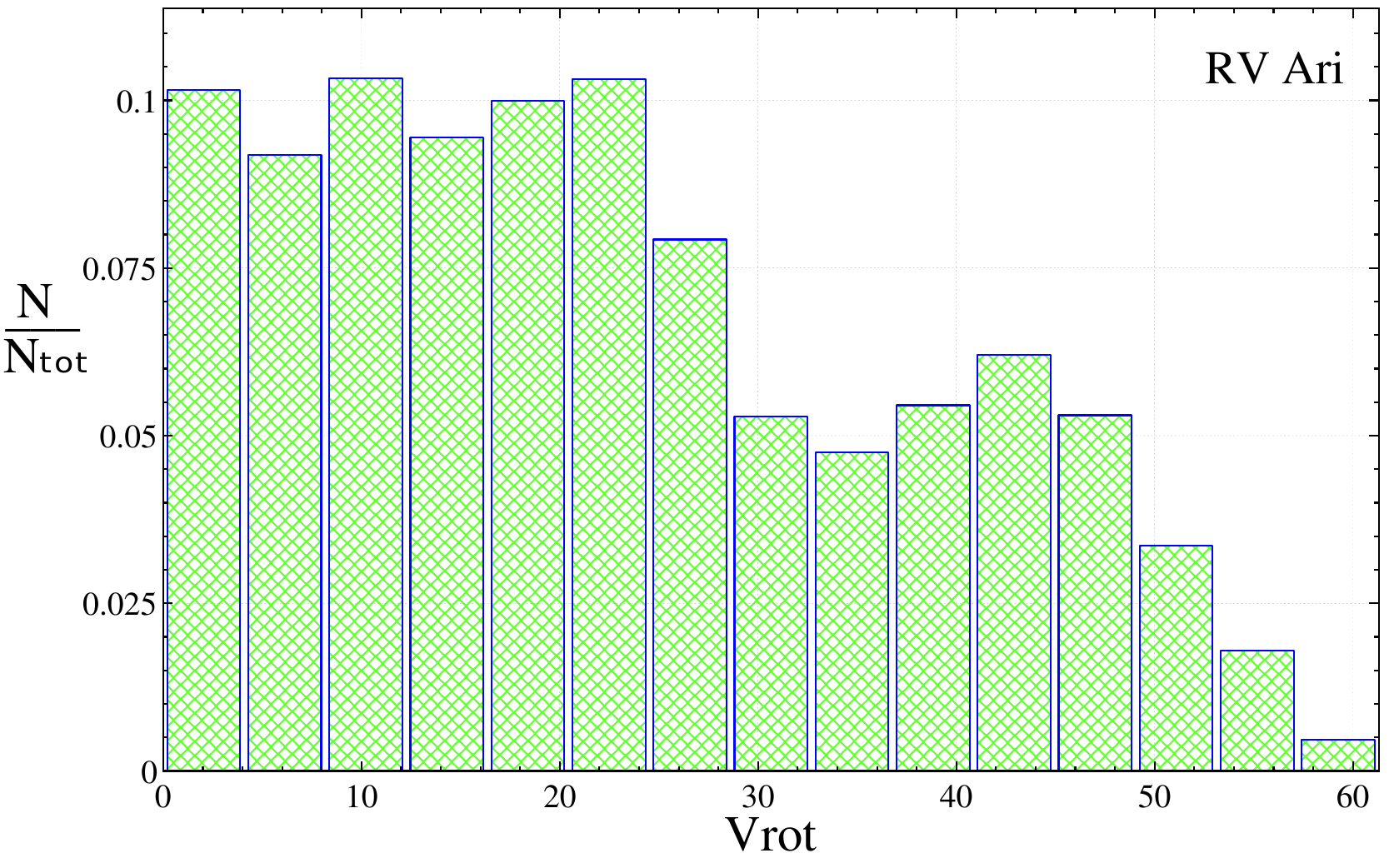}
	\includegraphics[clip,width=0.43\linewidth,height=58mm]{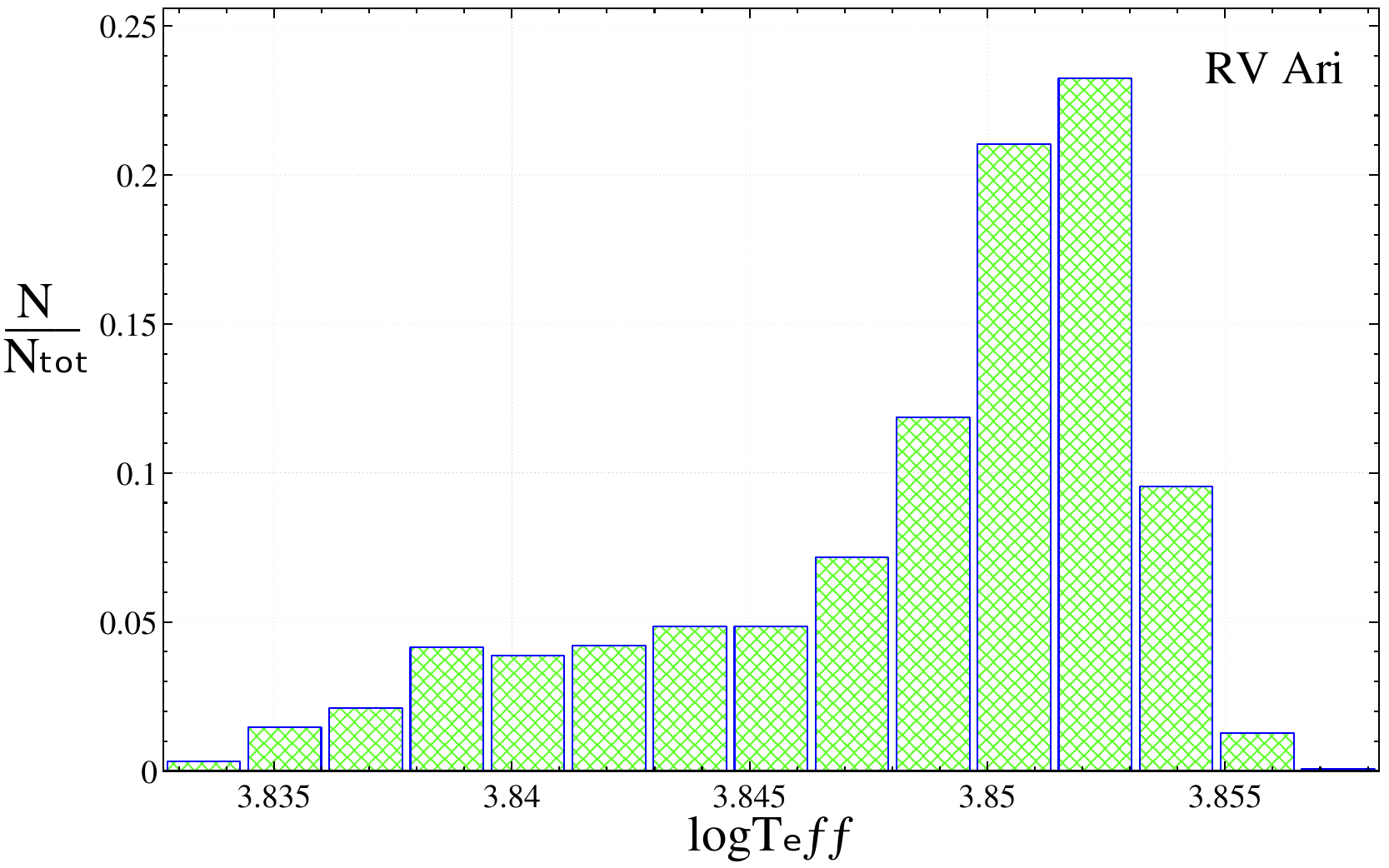}
	\includegraphics[clip,width=0.43\linewidth,height=58mm]{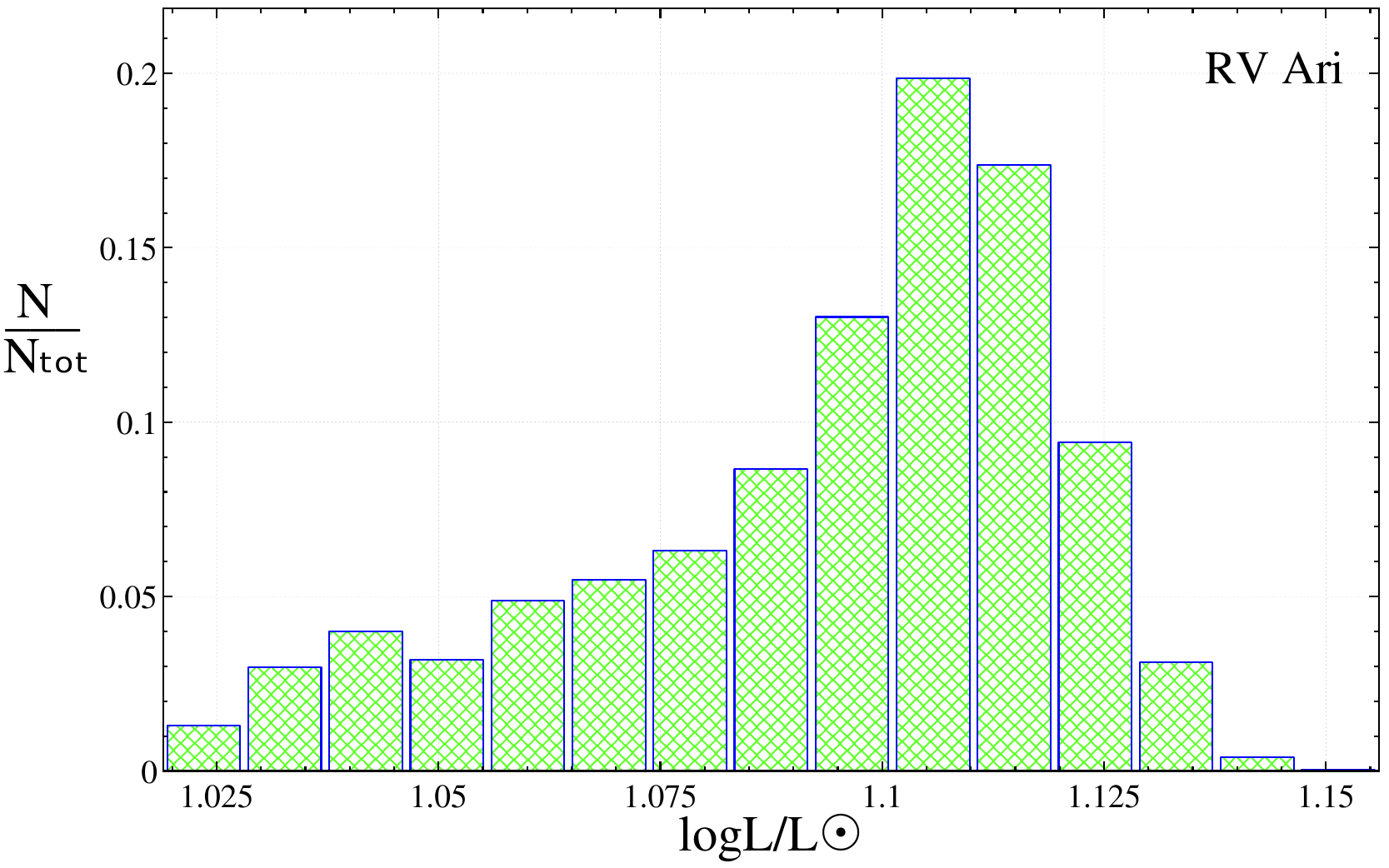}
	\includegraphics[clip,width=0.43\linewidth,height=58mm]{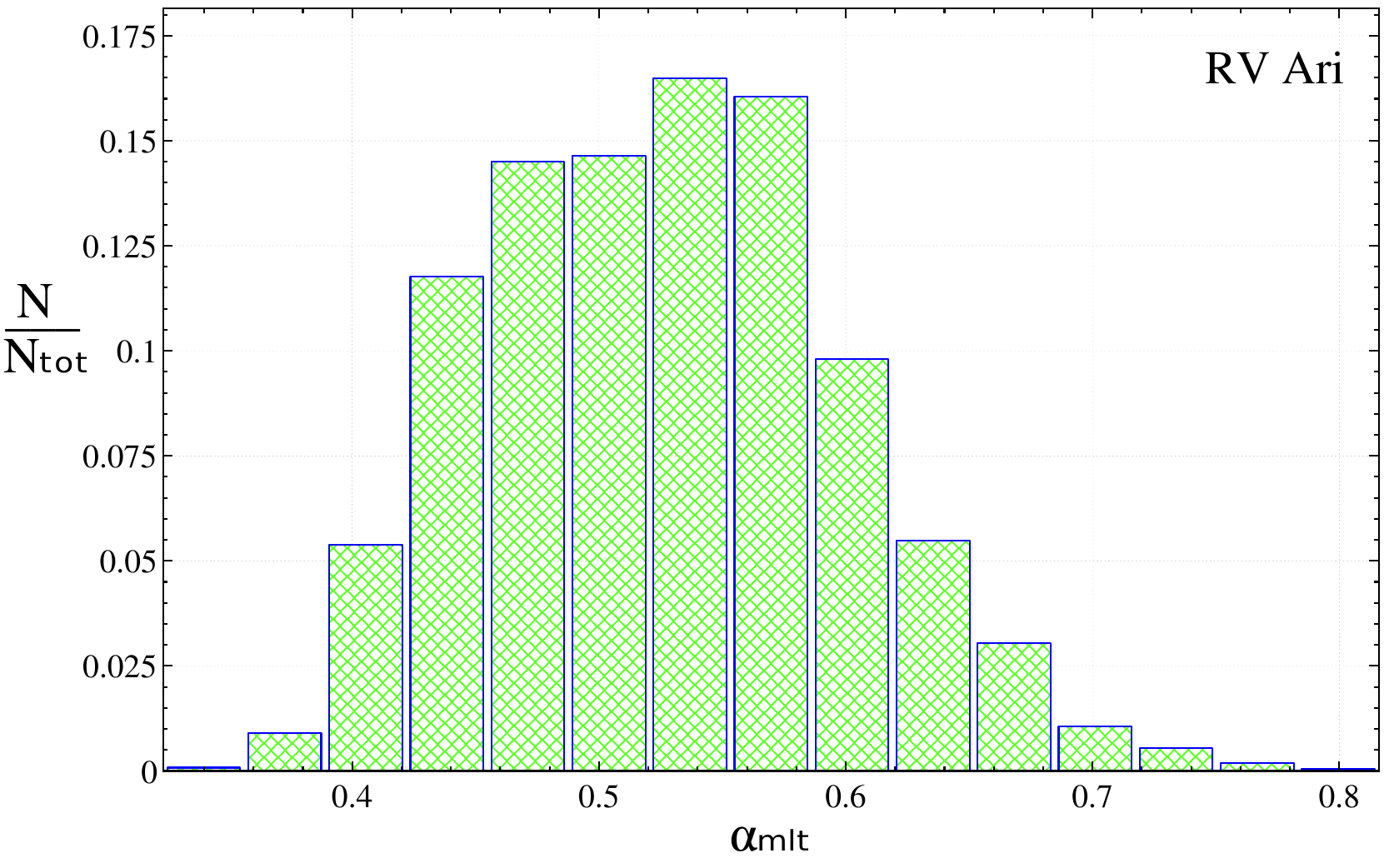}
	\includegraphics[clip,width=0.43\linewidth,height=58mm]{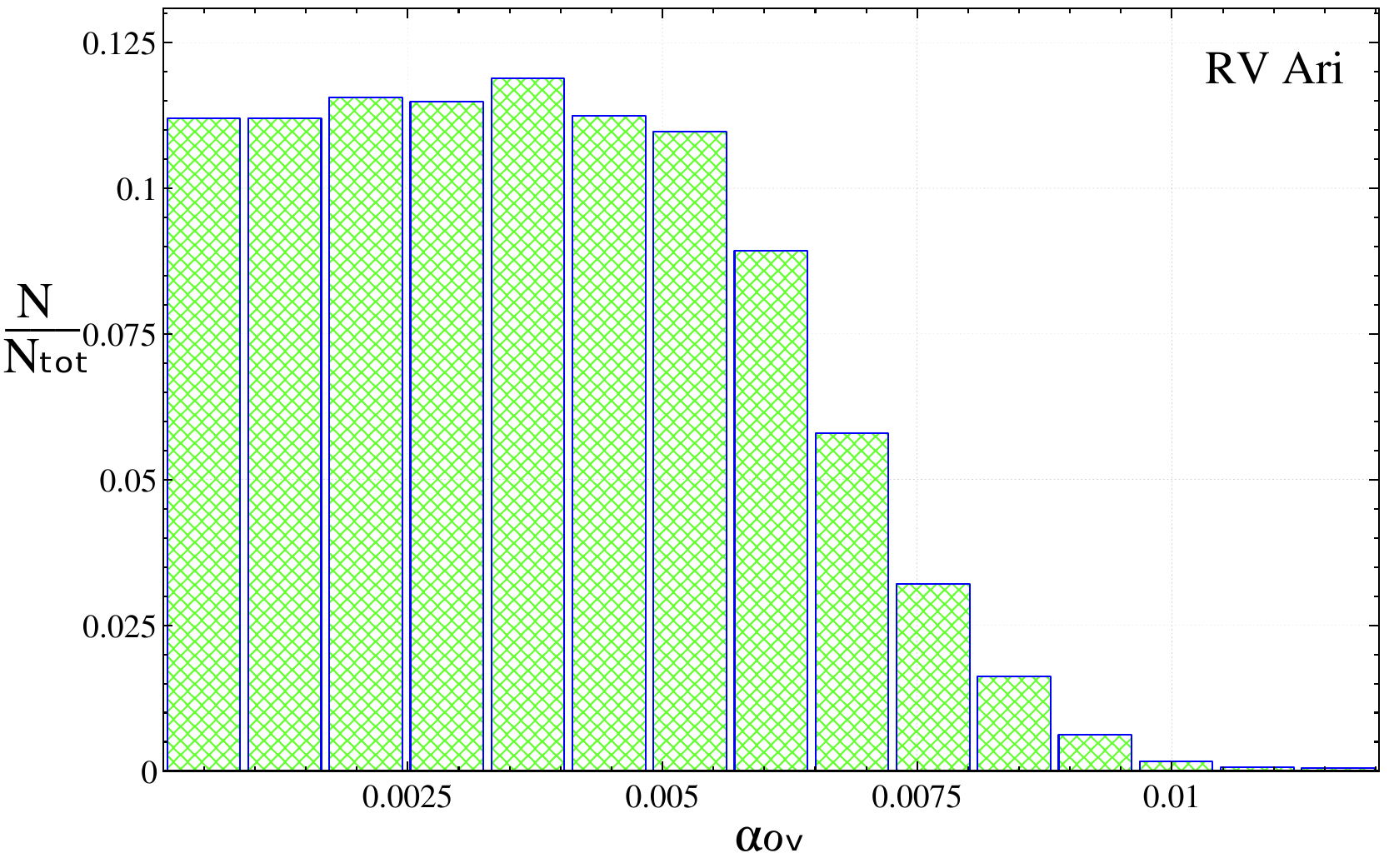}
	\caption{The corresponding histograms for determined parameters of RV Ari.}
\end{figure*}

\begin{figure*}
	\centering
	\includegraphics[clip,width=1.02\linewidth,height=120mm]{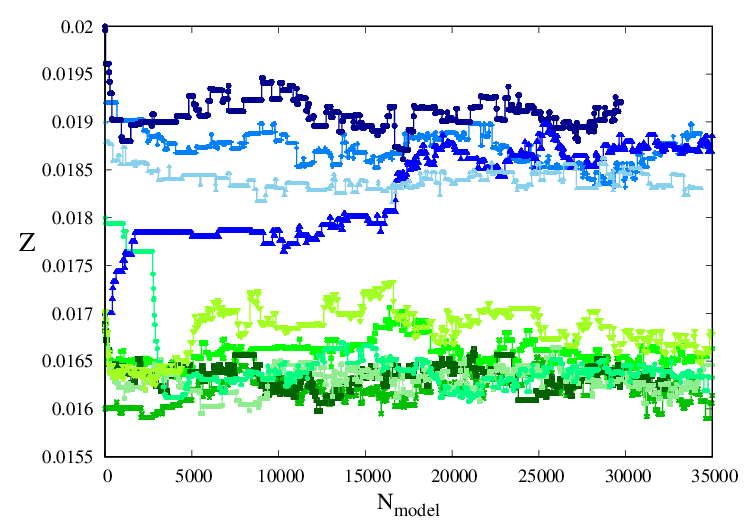}
	\caption{The metallicity $Z$ as a function of the model number for HSB models of RV Ari obtained from 
		the Bayesian analysis,  which aims to also fit the third frequency $\nu_3$ as a dipole axisymmetric mode.
		Bluish and greenish colours correspond to simulations converging to the higher and lower $Z$, respectively.}
\end{figure*}

\bsp	
\label{lastpage}
\end{document}